\documentclass[aps,prl,twocolumn,superscriptaddress,floatfix,showpacs]{revtex4-2}
\usepackage{graphicx}
\usepackage{amsmath,amssymb,amsfonts}
\usepackage{xcolor}
\usepackage{bm}
\usepackage{booktabs}
\usepackage[colorlinks=true,linkcolor=blue,citecolor=blue,urlcolor=blue]{hyperref}
\usepackage{threeparttable}
\usepackage{ragged2e}

\begin{document}

\title{Deformation-Potential-Driven Photostriction in Layered Ferroelectrics}

\author{S. Puri}
\affiliation{Department of Physics, University of Arkansas, Fayetteville, Arkansas, 72701, USA.}

\author{R. Rodriguez}
\affiliation{Department of Physics, University of Arkansas, Fayetteville, Arkansas, 72701, USA.}

\author{C. Dansou}
\affiliation{Department of Physics, University of Arkansas, Fayetteville, Arkansas, 72701, USA.}
\affiliation{Smart Ferroic Materials Center and Institute for Nanosciences \& Engineering, University of Arkansas, Fayetteville, 72701, Arkansas, USA.}

\author{L. Bouric}
\affiliation{Department of Physics, University of Arkansas, Fayetteville, Arkansas, 72701, USA.}
\affiliation{Smart Ferroic Materials Center and Institute for Nanosciences \& Engineering, University of Arkansas, Fayetteville, 72701, Arkansas, USA.}

\author{A. Sheibani}
\affiliation{Department of Physics, University of Arkansas, Fayetteville, Arkansas, 72701, USA.}

\author{C. Paillard}
\affiliation{Department of Physics, University of Arkansas, Fayetteville, Arkansas, 72701, USA.}
\affiliation{Smart Ferroic Materials Center and Institute for Nanosciences \& Engineering, University of Arkansas, Fayetteville, 72701, Arkansas, USA.}
\affiliation{Université Paris-Saclay, CentraleSupélec, CNRS, Laboratoire SPMS,  91190, Gif-sur-Yvette, France.}

\author{L. Bellaiche}
\affiliation{Department of Physics, University of Arkansas, Fayetteville, Arkansas, 72701, USA.}
\affiliation{Smart Ferroic Materials Center and Institute for Nanosciences \& Engineering, University of Arkansas, Fayetteville, 72701, Arkansas, USA.}
\affiliation{Department of Materials Science and Engineering, Tel Aviv University, Ramat Aviv, Tel Aviv 6997801, Israel.}

\author{H. Nakamura}
\email{hnakamur@uark.edu}
\affiliation{Department of Physics, University of Arkansas, Fayetteville, Arkansas, 72701, USA.}
\affiliation{Smart Ferroic Materials Center and Institute for Nanosciences \& Engineering, University of Arkansas, Fayetteville, 72701, Arkansas, USA.}

\date{\today}

\begin{abstract}
\noindent{}The coupling between electronic excitations and lattice deformation in van der Waals ferroelectrics is governed by a competition between the electron deformation potential and the inverse piezoelectric effect. While theory predicts that piezoelectric screening should drive a polar-axis contraction in monolayer group-IV monochalcogenides, we demonstrate that in multilayer SnS, the deformation potential provides the dominant contribution, driving a polar-axis expansion even within ferroelectric domains. By correlating polarization-resolved second-harmonic generation microscopy with ultrafast reflectance spectroscopy and first-principles calculations, we resolve the anisotropic lattice response and disentangle intrinsic photostrictive strain from extrinsic thin-film interference artifacts. These results establish a microscopic hierarchy of photostrictive mechanisms and position stacking-engineered SnS as a platform for ultrafast optomechanical transduction.
\end{abstract}

\maketitle
Layered group-IV monochalcogenides ($MX$; $M=\mathrm{Ge}, \mathrm{Sn}$; $X=\mathrm{S}, \mathrm{Se}$) have established themselves as a fertile ground for exploring the intersection of ferroelectricity and optoelectronics. Distinguished by a puckered crystal structure that mimics black phosphorus, these materials exhibit a robust in-plane spontaneous polarization \cite{wu2016intrinsic,wang2017two,bao2019gate,higashitarumizu2020purely} and strong optical anisotropy  \cite{zhu2021efficient,banai2016review,maragkakis2022nonlinear,sarkar2023liquid}. While their static properties---such as high carrier mobility \cite{xin2016few} and tunable band gaps \cite{huang2017layer}---are well-documented, the coupling between their electronic excitations and lattice degrees of freedom on ultrafast timescales stands as a key unresolved question. Specifically, the mechanisms governing photostriction (light-induced mechanical deformation) in these van der Waals ferroelectrics are debated, with two distinct forces often in competition: the electron deformation potential (DP) and the inverse piezoelectric effect (IPE).

\begin{figure}[htb]
    \centering
    \includegraphics[width=0.48\textwidth]{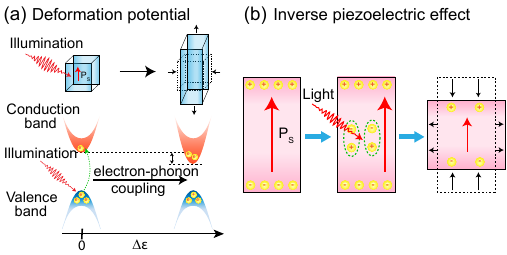}
    \caption{The two key mechanisms for photostriction. (a) Deformation-potential-driven photostriction, where optical excitation redistributes electronic population into excited states whose strain-dependent energies exert an internal electronic stress on the lattice. (b) Inverse piezoelectric contribution to photostriction, where illumination of a ferroelectric with spontaneous polarization (P$_s$) generates photocharges that partially screen the polarization, modifying the internal electric field and inducing crystal expansion/compression.}
    \label{fig:1_mechanism}
\end{figure}
The deformation potential mechanism describes how photoexcited carriers modify band-edge energies, creating an electronic stress that displaces the equilibrium lattice coordinates~\cite{thomsen1986surface, zeiger1992theory}---a force distinct from thermal expansion that drives the anisotropic structural changes [Fig.~\ref{fig:1_mechanism}(a)]. In contrast, in ferroelectric systems, the inverse piezoelectric effect often dictates an opposing response; photo-generated charges screen spontaneous polarization, reducing the internal electric field and inducing crystal compression [Fig.~\ref{fig:1_mechanism}(b)]~\cite{daranciang2012ultrafast,haleoot2017photostrictive,chen2025constructing}. This screening-driven contraction is a well-established phenomenon in ferroelectrics, having been observed in prototypical systems such as PbTiO$_3$ thin films~\cite{daranciang2012ultrafast} and polar oxides~\cite{chen2025constructing}. Seminal theoretical work on monolayer SnS similarly predicted that this inverse piezoelectric effect would dominate, driving a contraction of the polar axis upon illumination~\cite{haleoot2017photostrictive}. However, this picture is complicated by recent studies on GeS bulk crystals~\cite{luo2023ultrafast}, where the interplay between these mechanisms is inferred: experiments reveal expansion along the zigzag (ZZ) direction but compression along the polar armchair (AC) axis, consistent with the inverse piezoelectric prediction, while theoretical calculations for the same system suggest that deformation potential effects lead to a transient AC expansion for short timescales~\cite{luo2023ultrafast}. Resolving how these competing forces manifest requires a material platform where stacking orders and domain structures can be precisely interrogated.

In this Letter, we investigate multilayer SnS microcrystals and establish the dominant mechanism governing photostriction in this layered ferroelectric system. Using polarization-resolved second-harmonic generation (SHG), we first map the rich landscape of stacking orders, identifying the existence of ferroelectric (AA, AC) and antiferroelectric (AB) domains among distinct SnS multilayers or even coexistence within an individual microcrystal. We then utilize ultrafast reflectance spectroscopy to reveal a pronounced, domain-dependent modulation of the refractive index. Strikingly, we observe a sign reversal in the transient signal between orthogonal crystallographic axes---a behavior we attribute to anisotropic photostriction where the deformation potential overcomes inverse piezoelectric screening. This drives polar-axis expansion even in ferroelectric multilayers, where a sizable inverse piezoelectric response would typically mandate contraction. By combining these observations with a thin-film interference model, we further demonstrate an \textit{apparent} sign reversal of reflectance modulations for the same crystallographic direction at different thicknesses, which originates purely from the interplay between the refractive index change and multiple optical reflections. These results establish SnS not just as a ferroelectric semiconductor but also as a tunable platform for manipulating light-matter interactions through structural design.

\begin{figure}[htb]
    \centering
    \includegraphics[width=0.48\textwidth]{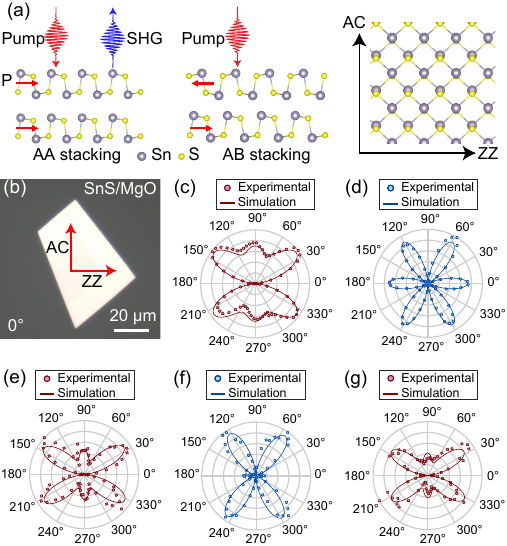}
    \caption{Anisotropy in SHG generated by stacking. (a)(Left) Schematics showing the presence of SHG in AA(FE) stacking and its absence in AB(AFE) stacking. (Right) Atomic structure of SnS (top-view) indicating armchair (AC) and zigzag (ZZ) directions. (b) Optical image of monodomain SnS with thickness of 49.5 nm on MgO substrate.   (c-g) Polarization-resolved SHG intensity plot for (c,d) 49.5 nm thick monodomain SnS on MgO substrate under (c) parallel, and (d) perpendicular configuration.
    (e,f) 7.7 nm thick monodomain SnS on mica substrate under (e) parallel, and (f) perpendicular configuration.
    (g) 65.5 nm thick monodomain SnS on mica substrate under parallel configuration. Solid line represents the simulated fit using the two-tensor model. }
    \label{fig:2_shg_anisotropy}
\end{figure}

We first examined the nonlinear optical response of PVD-grown SnS microcrystals using polarization-resolved SHG to identify regions with non-centrosymmetric stacking symmetry. Strong SHG signals were observed in many islands, indicating the presence of ferroelectric (FE) layer-by-layer stacking sequences, as shown schematically in [Fig.~\ref{fig:2_shg_anisotropy}(a)]. The expected quadratic dependence of SHG intensity on excitation power confirms the second-order origin of the signal [Fig. S1~\cite{supmat}].

The stacking configuration dictates the symmetry of the SHG response. Centrosymmetric AB-stacked SnS belongs to the space group \textit{Pnma} (No.\,62) and point group $D_{2h}$~\cite{bao2019gate,wang2017giant}, for which all $\chi^{(2)}$ elements vanish. In contrast, the ferroelectric AA ($Pmn2_1$) and AC ($Pmm2$) stacking motifs share the non-centrosymmetric point group $C_{2v}$, allowing three independent non-vanishing tensor components: $\chi^{(2)}_{yxx}$, $\chi^{(2)}_{xyx}$, and $\chi^{(2)}_{yyy}$~\cite{yonemori2021thickness,shi2016symmetry}. The expected angular dependence is given by~\cite{moqbel2024giant}:
\begin{equation}
I_{||}(2\omega) \propto (2\chi_{xyx}^{(2)} + \chi_{yxx}^{(2)})\sin\theta\cos^2\theta + \chi_{yyy}^{(2)}\sin^3\theta,
\end{equation}
\begin{equation}
I_{\perp}(2\omega) \propto \chi^{(2)}_{yxx}\cos^3\theta + (\chi^{(2)}_{yyy} - 2\chi^{(2)}_{xyx})\cos\theta\sin^2\theta,
\end{equation}
where $\theta$ is the pump polarization angle with respect to ZZ direction.

\begin{figure}[htb]
    \centering
    \includegraphics[width=0.48\textwidth]{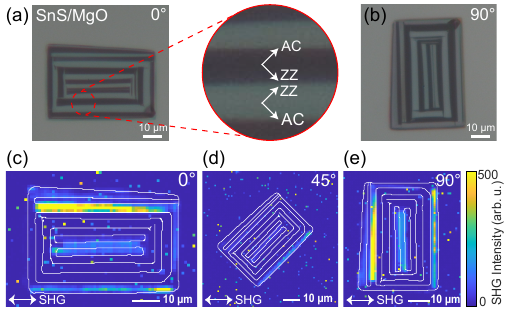}
    \caption{Coexistence of FE and AFE stacking in a SnS island. (a,b) Cross polarized microscopy image of 270 nm thick multidomained SnS on MgO substrate at (a) 0$^{\circ}$, and (b) 90$^{\circ}$ orientation.  Relative bright and dark stripes represent the different crystallographic domains. The magnified image shows the AC and ZZ directions for the bright and dark striped region for 0 $^{\circ}$ orientation of the sample. (c-e) SHG mapping of multidomained SnS island under parallel configuration -  with both pump and SHG polarizations kept horizontal - at (c) 0$^{\circ}$, (d) 45$^{\circ}$, and (e) 90$^{\circ}$ orientation.  Bright regions in the SHG map represent the SHG active region. The white solid lines represent the outline of the pattern extracted from cross polarized optical image of multidomained SnS. Scale bar represents 10 $\mu$$m$.}
    \label{fig:3_shg_mapping}
\end{figure}

\begin{table}[tb]
    \centering
    \caption{Relative contribution of \textit{Type I} and \textit{Type II} $\chi^{(2)}$ components and corresponding SHG intensity ratio (I$_{max}$/I$_{AC}$) for SnS islands of different thickness.
    \label{tab:chi2_contribution}}
    \renewcommand{\arraystretch}{1.5} % Adjust the number to your liking
   \resizebox{\columnwidth}{!}{%
    \begin{tabular}{llccc}
        \toprule
        \toprule
        Substrate & SnS thickness & \textit{Type I} & \textit{Type II} & Ratio\\
        & & & & $I_{max}/I_{AC}$\\ \toprule
        \toprule

        MgO & 49.5 nm & 53 $\%$ & 47 $\%$ & 1.44 \\
        Mica & 7.7 nm & 36 $\%$ & 64 $\%$ & 1.98 \\ 
        Mica & 65.5 nm & 39 $\%$ & 61 $\%$ & 1.93 \\ \bottomrule
        
    \end{tabular}%
    }
\end{table}

Fig.~\ref{fig:2_shg_anisotropy}(c) and Fig.~\ref{fig:2_shg_anisotropy}(d) displays the SHG polar pattern for a monodomain SnS island (49.5\,nm thick) on MgO [Fig.~\ref{fig:2_shg_anisotropy}(b)]. The data reveal a dominant two-lobed ``butterfly'' shape with a distinct central hump—a feature that cannot be captured by a single $C_{2v}$ tensor. We found that this complex shape is accurately reproduced by a two-tensor model, in which two distinct $\chi^{(2)}$ tensors with different magnitudes (and opposite signs for $\chi^{(2)}_{yyy}$) contribute to the total signal. These are defined as \textit{Type I} and \textit{Type II} tensors:

\[
\textit{Type I} : \chi^{(2)}_{\textit{Type I}} = 
\begin{pmatrix}
    0 & 0 & 0 & 0 & 0 & \chi^{(2)}_{xyx} \\
    \chi^{(2)}_{yxx} & \chi^{(2)}_{yyy} & 0 & 0 & 0 & 0 \\
    0 & 0 & 0 & 0 & 0 & 0
\end{pmatrix}
\]
 and
 \[
\textit{Type II} : \chi^{(2)}_{\textit{Type II}} = 
\begin{pmatrix}
    0 & 0 & 0 & 0 & 0 & \chi^{(2)}_{xyx} \\
    \chi^{(2)}_{yxx} & - \chi^{(2)}_{yyy} & 0 & 0 & 0 & 0 \\
    0 & 0 & 0 & 0 & 0 & 0
\end{pmatrix}
\]

This multi-tensor response suggests that the SHG-active regions are not monolithic but rather host distinct ferroelectric stacking domains (e.g., variations between AA and AC motifs) or stacking faults along the optical path. This behavior is robust across different substrates and thicknesses [Fig.~\ref{fig:2_shg_anisotropy}(e) - Fig.~\ref{fig:2_shg_anisotropy}(g)], with the relative contribution of the two tensors varying with film thickness [Table~\ref{tab:chi2_contribution}].  Notably,  the intensity along the AC direction (in parallel configuration) and the ZZ direction (in perpendicular configuration) increases when the contribution from the \textit{Type I} outweighs that of \textit{Type II}.

To spatially resolve these structural variations, we correlated cross-polarized optical microscopy with SHG mapping. Fig.~\ref{fig:3_shg_mapping}(a) and Fig.~\ref{fig:3_shg_mapping}(b) shows a multidomain SnS island exhibiting pronounced bright and dark stripe patterns under cross-polarization microscopy. Rotating the sample by $90^{\circ}$ reverses the contrast of these stripes, indicating that they correspond to crystallographic domains whose orientations differ by $90^{\circ}$~\cite{sutter2024macroscopic,mao2023giant}.

\begin{figure}[tb]
    \centering
    \includegraphics[width=0.48\textwidth]{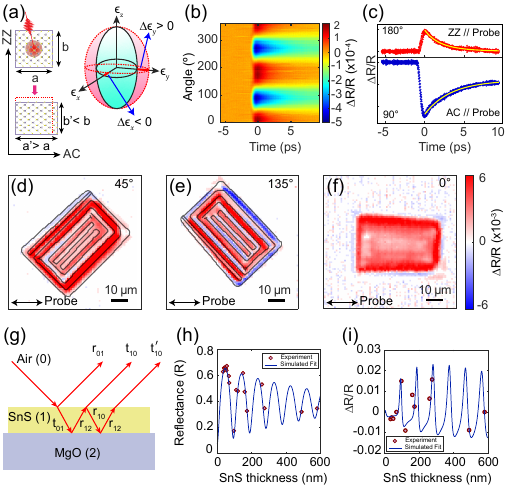}
    \caption{Sign reversal in transient reflection signal. (a) Schematics showing photostriction in SnS (left). Dielectric ellipsoid showing the change in dielectric constant due to photostriction (right). (b) Color map of polarization dependent transient reflection signal for 49.5 nm thick monodomain SnS. (c) Representative positive and negative reflective signal from the color map of (b) for 180$^{\circ}$ and 90$^{\circ}$, respectively. Blue and red circles represents the experimental data points whereas the solid yellow line represents the biexponential fit to the decay dynamics. (d-f) Color map of peak intensity of transient reflection signal from 270 nm thick multi-domain SnS for sample orientation at (d) 45$^{\circ}$, (e) 135$^{\circ}$, and (f) 0$^{\circ}$, while horizontal probe polarization direction was kept. Dark solid lines overlaid in (d) and (e) are a guide to identify the domains. Scale bar represents 10 $\mu$$m$.  (g) Multireflection diagram for SnS film on MgO substrate. (h) Thickness dependent reflectance taken with probe polarization parallel to AC. Red dots (blue solid curve) shows the experimental data (simulated fit). (i) Transient reflectivity change as a function of SnS film thickness, with probe polarization parallel to AC. Red dots represent the experimental data, and blue solid curve shows the simulated fit incorporating interference effects and index modulation from photostriction.}
    \label{fig:4_pp_mapping}
\end{figure}

SHG mapping of the same island [Fig.~\ref{fig:3_shg_mapping}(c) - Fig.~\ref{fig:3_shg_mapping}(e)] reveals that only specific regions exhibit nonlinear activity. Bright stripes in the SHG map align with crystallographic domains where non-centrosymmetric stacking is present, with intensity maximizing when the pump polarization aligns with the local lattice symmetry. Crucially, other regions of the island remain dark in the SHG map despite being visible in optical microscopy. This absence of SHG signal indicates regions of centrosymmetric stacking (likely pure AFE/AB stacking) devoid of significant ferroelectric stacking faults. Thus, in the subset of crystals displaying such domain patterns, we uncover a spatially heterogeneous architecture composed of interleaved SHG-active ferroelectric domains and SHG-inactive antiferroelectric regions. These results establish that PVD-grown SnS forms a complex, stacking-engineered platform suitable for investigating the domain-dependent ultrafast dynamics discussed below.

To probe how these structural motifs influence ultrafast dynamics, we performed degenerate pump--probe reflectance spectroscopy. Fig.~\ref{fig:4_pp_mapping}(b) shows the polarization-resolved transient reflectance $\Delta R/R_0$ for the 49.5\,nm monodomain island. As the sample is rotated, the transient signal exhibits a striking $\pi$-periodic modulation: the ZZ direction yields a positive reflectivity change, whereas the AC direction shows a negative one. This strongly anisotropic, sign-opposite response cannot be accounted for by isotropic thermal expansion, which would produce a uniform reflectance change independent of crystallographic orientation. Instead, this behavior stems from pump-induced lattice deformation, which results in an anisotropic dielectric change along the crystallographic directions, as schematically shown in Fig.~\ref{fig:4_pp_mapping}(a).

Representative transient reflectance traces for the AC and ZZ axes are shown in Fig.~\ref{fig:4_pp_mapping}(c). Both signals are well described by a biexponential decay,
\begin{equation}
\Delta R(t)/R_0 = A_1 e^{-t/\tau_1} + A_2 e^{-t/\tau_2},
\end{equation}
where $\tau_1$ captures the fast electronic relaxation associated with carrier recombination, and $\tau_2$ describes a slower relaxation process. Notably, the AC direction exhibits a substantially longer $\tau_2$ (25.99 $\pm$ 7.47\,ps) compared to the ZZ direction (8.12 $\pm$ 2.96\,ps), indicating a stronger coupling to lattice deformation along the polar axis. The slower component $\tau_2$, which is strongly anisotropic and enhanced along the polar axis, is therefore assigned to lattice deformation rather than carrier recombination.

To assess whether this anisotropy is merely a selection-rule effect governed by a relationship between the polarization direction of a pump beam with respect to the crystallographic axes, we performed pump-probe measurements on a 275\,nm thick monodomain SnS island with pump polarization oriented either parallel or perpendicular to the probe [Fig. S2~\cite{supmat}]. In both configurations, the sign of the transient signal remained unchanged, demonstrating that the response is governed by the orientation of the probe polarization with respect to crystallographic axis rather than the pump polarization. The response amplitude, however, is found to be larger in the parallel configuration.

A related question is whether the ultrafast reflectance modulation is governed by crystallographic domains or by ferroelectric polarization domains. To address this, we performed spatial mapping of the pump--probe response on a multidomained SnS island. Fig.~\ref{fig:4_pp_mapping}(d - f) show color maps of the peak $\Delta R/R_{0}$ at several sample orientations while keeping the probe polarization fixed. The mapped response exhibits a stripe-like contrast that rotates with the sample and switches between neighboring regions with a $\sim 90^{\circ}$ periodicity, consistent with domains whose principal crystallographic axes are rotated by approximately $90^{\circ}$ relative to one another. Importantly, the boundaries and contrast of the transient maps coincide with the crystallographic domain pattern identified optically, indicating that the anisotropic reflectance modulation is governed primarily by crystallographic orientation rather than by ferroelectric polarization domain contrast. This behavior was confirmed also in another multidomained SnS with stripe-domains identified by the polarization microscopy [Fig. S3~\cite{supmat}], by comparing spatially-resolved reflectance and SHG mappings [Fig. S4 and Fig. S5~\cite{supmat}].

While the sign reversal between AC and ZZ axes is striking, the raw sign of $\Delta R/R_0$ in thin films is often obscured by optical interference as schematically shown in Fig. ~\ref{fig:4_pp_mapping}(g). To disentangle intrinsic material responses from extrinsic geometric effects, we measured the static and transient reflectance across a range of SnS thicknesses using islands that showed strong SHG. Fig.~\ref{fig:4_pp_mapping}(h) shows that the static reflectance exhibits clear oscillatory fringes. Fitting these curves following the optical interference model (see Section S8~\cite{supmat}) yields complex refractive indices of $4.23 + 0.12i$ (AC) and $4.36 + 0.245i$ (ZZ), in agreement with literature trends~\cite{banai2016review}.

Crucially, the transient signal $\Delta R/R_0$ also undergoes a thickness-dependent sign reversal. For instance, the 49.5\,nm film shows $\Delta R/R_0 > 0$ along ZZ and $< 0$ along AC [Fig.~\ref{fig:4_pp_mapping}(b)], whereas a 275\,nm film shows the exact opposite behavior [Fig. S2(d)~\cite{supmat}]. We modeled this response using a standard multilayer interference formalism (see Section S8~\cite{supmat}), introducing the pump-induced perturbation as small changes in the complex refractive index ($\Delta n$ and $\Delta k$). The calculated curves closely match the measured thickness dependence [Fig.~\ref{fig:4_pp_mapping}(i)], confirming that the observed sign flips are interference-driven artifacts. This analysis provides a general protocol for extracting intrinsic photostrictive strain from pump--probe measurements in van der Waals multilayers.

Extracting the intrinsic response from this model yields $\Delta n = -0.00302$ $\pm$ 0.00086 along AC and $\Delta n = +0.00022$ $\pm$ 0.00067 along ZZ. The sign of these refractive index changes is consistent with a prior study on a bulk SnS single crystal~\cite{sun2022dynamical}. Negative $\Delta n$ along the polar AC axis implies a lattice expansion, while positive $\Delta n$ along ZZ implies contraction. This result is pivotal: theoretical predictions for monolayer SnS based on the inverse piezoelectric effect (IPE) predict a \textit{contraction} of the polar axis due to the screening of spontaneous polarization~\cite{haleoot2017photostrictive}. The observed expansion therefore indicates that the IPE is overwhelmed by a competing mechanism, even in multilayers that exhibit ferroelectric stacking. Furthermore, quantitative estimation of the pump-induced temperature rise yeilds a maximum thermal strain($\sim10^{-6}$) that is at least two orders of magnitude too small to account for the structural strains ($\sim10^{-4}$) required for these refractive index modulations (see section S9~\cite{supmat}).  This negligible magnitude, combined with the inherently isotropic nature of thermal expansion~\cite{luo2023ultrafast},  unequivocally rules out simple heating as the origin of the observed highly anisotropic, sign-opposite lattice deformation.

We attribute this behavior to the electron deformation potential (DP), where photoexcited carriers modify band-edge bonding forces to generate an internal stress that drives expansion along the polar axis [Fig.~\ref{fig:1_mechanism}(a)]. This interpretation aligns with studies on bulk ferroelectrics such as BaTiO$_3$~\cite{hoang2025ultrafast} and BiFeO$_3$~\cite{wen2013electronic}, where deformation potential effects have been shown to compete with or dominate the inverse piezoelectric contraction (Table S1 \cite{supmat}). We note that this monotonic expansion contrasts with the complex temporal response observed in out-of-plane ferroelectric thin films, such as PbTiO$_3$, which exhibit an initial contraction followed by expansion~\cite{daranciang2012ultrafast}. In such systems, the dynamics are driven by the screening of a strong out-of-plane depolarization field. In SnS, however, the in-plane orientation of the polarization minimizes the impact of the depolarization field due to the geometry, thereby simplifying the response and allowing the deformation potential to dominate. Thus, even in SnS multilayers where a sizable inverse piezoelectric response is expected, the deformation potential stress is sufficient to dictate the macroscopic structural response.

\begin{figure}[htb]
    \centering
    \includegraphics[width=0.48\textwidth]{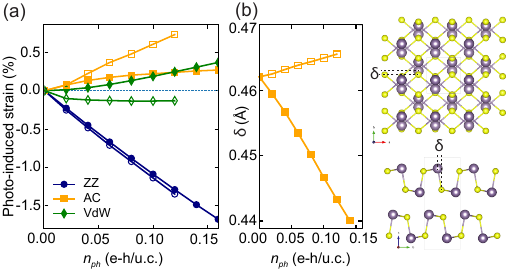}
    \caption{DFT-calculated photostriction in AFE SnS. (a) The relative change in lattice constants along the AC (armchair) and ZZ (zigzag) and out-of-plane (VdW) axes as a function of photo-excited carrier concentration ($n_{ph}$) for bulk SnS with antiferroelectric stacking. Filled symbol represent calculations in which all ionic degrees of freedom are allowed to relax, meanwhile open symbols indicate that the reduced coordinates of ions were frozen in the AC direction during relaxation, preventing electric dipole reduction. These latter calculations isolate the deformation potential contribution, predicting AC expansion and ZZ contraction, consistent with experimental observations. (b) DFT-calculated evolution of the spacing $\delta$ between a Sn and S atom under photoexcitation, qualitatively showing the change in local dipoles when all ions are relaxed (filled symbols) or their reduced coordinate frozen along the AC axes (open symbols).}
    \label{fig:5_dft}
\end{figure}

To corroborate this mechanism, we performed density functional theory (DFT) calculations on bulk SnS (see Supplementary Materials~\cite{supmat}). As shown in Fig.~\ref{fig:5_dft}(a), the calculations reveal a striking anisotropy in the lattice response: increasing the excited carrier concentration triggers a monotonic expansion along AC axis and a contraction along ZZ axis. For a carrier concentration of $n_{ph} = 0.1$ e-h/u.c., the lattice expands by approximately 0.2\% along the AC direction, while ZZ direction contracts by 1\%. Crucially, this theoretical prediction for the AFE phase matches the sign of the intrinsic strain we extracted experimentally. Estimating the lateral offset between Sn and S atoms in the AC direction (a qualitative estimate of the local dipoles) in Fig.~\ref{fig:5_dft}(b) indicates that, as in most thick ferroelectric or antiferroelectrics ~\cite{paillard2016photostriction,paillard2017ab,dansou2025photoinduced}, local dipoles decrease under photoexcitation. If the primary mechanism of photo-induced deformation was a reduction in local dipoles associated with a local inverse piezoelectric effect (note that such a mechanism is also allowed in antiferroelectrics ~\cite{dansou2025photoinduced}, even though they are not macroscopically piezoelectric), one would expect a contraction of the AC axis. In contrast, our calculation reveals an expansion of the AC axis. As a further confirmation, we have performed calculations freezing the internal coordinates of the atoms along the AC direction, approximately freezing local dipoles (in fact, a small increase is observed in Fig.~\ref{fig:5_dft}(b)) and mitigating inverse piezoelectric effects. Calculating the photo-induced deformation in these conditions shows in Fig.~\ref{fig:5_dft}(a) that the AC expansion is even stronger, meanwhile ZZ contraction remains quantitatively similar.
The DFT calculations independently reproduce the experimentally extracted anisotropic strain, demonstrating that deformation-potential stress alone is sufficient to generate polar-axis expansion in SnS.

We have resolved the competing mechanisms governing photostriction in layered SnS, demonstrating that the deformation potential provides the dominant contribution to ultrafast lattice modulation, overriding inverse piezoelectric screening even within ferroelectric stacking domains. By correlating polarization-resolved second-harmonic generation microscopy with ultrafast pump--probe reflectance spectroscopy, we revealed a complex structural landscape in which ferroelectric and antiferroelectric stacking motifs coexist, yet exhibit a common intrinsic lattice response characterized by polar-axis expansion driven by electronic excitation. Through a combined experimental and theoretical analysis, we further established that this anisotropic strain response is an intrinsic property of SnS, while apparent sign reversals in transient reflectance originate from thin-film optical interference.

Beyond resolving a long-standing ambiguity in the photostriction of group-IV monochalcogenides, our results establish stacking-engineered SnS as a versatile platform for ultrafast optomechanical transduction, in which electronic excitation can be efficiently converted into directional lattice motion. More broadly, the demonstrated dominance of deformation-potential-driven strain suggests a general design principle for photo-driven ferroelectric and ferroelastic devices, enabling light-controlled actuation without relying on polarization screening alone. These findings open new opportunities for engineering ultrafast, anisotropic mechanical responses in van der Waals (anti)ferroelectrics and related heterostructures through control of stacking symmetry, domain architecture, and electronic excitation.

\begin{acknowledgments}
This work was supported by the Office of the Secretary of Defense for Research and Engineering (Award No. FA9550-23-1-0500), Air Force Office of Scientific Research (Award No. FA9550-24-1-0263), and the National Science Foundation through the MonArk NSF Quantum Foundry (Award No. DMR-1906383).
\end{acknowledgments}

%\bibliography{SnS}


%%%%%%%%%% Merge with supplemental materials %%%%%%%%%%
\clearpage
\onecolumngrid
%\appendix

% Reset counters for figures, tables, etc.
\setcounter{figure}{0}
\setcounter{table}{0}
\setcounter{equation}{0}
\setcounter{page}{1}
\setcounter{secnumdepth}{2} % number sections + subsections (use 3 for subsubsections)

% Prefix "S" to labels
\renewcommand{\thesection}{S\arabic{section}}
\renewcommand{\thesubsection}{S\arabic{section}.\arabic{subsection}}
\renewcommand{\thefigure}{S\arabic{figure}}
\renewcommand{\thetable}{S\arabic{table}}
\renewcommand{\theequation}{S\arabic{equation}}

% Prefix "S" to hyperref internal identifiers to avoid duplicate destination warnings
\renewcommand{\theHsection}{S\arabic{section}}
\renewcommand{\theHsubsection}{S\arabic{section}.\arabic{subsection}}
\renewcommand{\theHfigure}{S\arabic{figure}}
\renewcommand{\theHtable}{S\arabic{table}}
\renewcommand{\theHequation}{S\arabic{equation}}

%%%%%%%%%% End Merge setup %%%%%%%%%%

% --- SM title block with correct affiliations ---
\begin{center}
\vspace*{-1.0em}
\begin{minipage}{0.92\textwidth}
\centering

{\large\bfseries Supplemental Material for}\\[2pt]
{\large\bfseries Deformation-Potential-Driven Photostriction in Layered Ferroelectrics}\\[8pt]

{\normalsize
S. Puri\textsuperscript{1},
R. Rodriguez\textsuperscript{1},
C. Dansou\textsuperscript{1,2},
L. Bouric\textsuperscript{1,2},
A. Sheibani\textsuperscript{1},
C. Paillard\textsuperscript{1,2,3},
L. Bellaiche\textsuperscript{1,2,4},
H. Nakamura\textsuperscript{1,2}\\[4pt]

\textsuperscript{1}\textit{Department of Physics, University of Arkansas, Fayetteville, AR 72701, USA}\\
\textsuperscript{2}\textit{Smart Ferroic Materials Center and Institute for Nanosciences \& Engineering, University of Arkansas, Fayetteville, 72701, Arkansas, USA.}\\
\textsuperscript{3}\textit{Université Paris-Saclay, CentraleSupélec, CNRS, Laboratoire SPMS,  91190, Gif-sur-Yvette, France.}\\
\textsuperscript{4}\textit{Department of Materials Science and Engineering, Tel Aviv University, Ramat Aviv, Tel Aviv 6997801, Israel.}\\

%\href{mailto:hnakamur@uark.edu}{hnakamur@uark.edu}
}

\end{minipage}
\vspace*{0.5em}
\end{center}
% --- end SM title block ---

% ---------------------------------------------------------
% SECTIONS
% ---------------------------------------------------------

% Table

\begin{table*}[!h]
\centering
\caption{Summary of photostriction mechanisms and polar-axis expansion.}
\label{tab:summary_table}

\renewcommand{\arraystretch}{2} % Adjust the number to your liking

\begin{threeparttable}
\begin{tabular}{p{2.6cm}p{2cm}p{2.6cm}p{3cm}p{3cm}c}
\hline
\hline
\parbox[t]{2.6cm}{\raggedright \textbf{Material}} &
\parbox[t]{2cm}{\raggedright \textbf{Method}} &
\parbox[t]{2.6cm}{\raggedright \textbf{Polar axis (PA)}} &
\parbox[t]{3cm}{\raggedright \textbf{PA expansion}} &
\parbox[t]{3cm}{\raggedright \textbf{Mechanism}} &
\textbf{Ref.} \\
\hline
\hline

\parbox[t]{2.6cm}{\raggedright \textbf{SnS multilayer}} &
\parbox[t]{2cm}{\raggedright \textbf{This work}} &
\parbox[t]{2.6cm}{\raggedright \textbf{AC (FE/AFE)}} &
\parbox[t]{3cm}{\raggedright \textbf{Yes}} &
\parbox[t]{3cm}{\raggedright \textbf{Deformation potential (DP)}} &
\textbf{--} \\

\parbox[t]{2.6cm}{\raggedright SnS \& SnSe (ML)} &
\parbox[t]{2cm}{\raggedright Theory} &
\parbox[t]{2.6cm}{\raggedright AC (FE)} &
\parbox[t]{3cm}{\raggedright No (contracts)} &
\parbox[t]{3cm}{\raggedright IPE} &
\cite{SM_haleoot2017photostrictive} \\

\parbox[t]{2.6cm}{\raggedright GeS bulk} &
\parbox[t]{2cm}{\raggedright Exp.+Th.} &
\parbox[t]{2.6cm}{\raggedright AC (AFE)} &
\parbox[t]{3cm}{\raggedright No (Exp.) Yes (Th.)} &
\parbox[t]{3cm}{\raggedright DP competes with IPE} &
\cite{SM_luo2023ultrafast} \\

\parbox[t]{2.6cm}{\raggedright BiFeO$_3$} &
\parbox[t]{2cm}{\raggedright Exp.} &
\parbox[t]{2.6cm}{\raggedright [111] (FE)} &
\parbox[t]{3cm}{\raggedright unmeasured} &
\parbox[t]{3cm}{\raggedright PVE + IPE} &
\cite{SM_zhang2025giant,SM_wen2013electronic,SM_kundys2010light} \\

\parbox[t]{2.6cm}{\raggedright PbTiO$_3$} &
\parbox[t]{2cm}{\raggedright Exp.} &
\parbox[t]{2.6cm}{\raggedright c-axis (FE)} &
\parbox[t]{3cm}{\raggedright Yes (contracts then expands)} &
\parbox[t]{3cm}{\raggedright PVE + IPE} &
\cite{SM_daranciang2012ultrafast} \\

\parbox[t]{2.6cm}{\raggedright BaTiO$_3$} &
\parbox[t]{2cm}{\raggedright Exp.+Th.} &
\parbox[t]{2.6cm}{\raggedright c-axis (FE)} &
\parbox[t]{3cm}{\raggedright Yes (contracts then expands)} &
\parbox[t]{3cm}{\raggedright IPE + DP} &
\cite{SM_hoang2025ultrafast} \\

\hline
\hline
\end{tabular}

\vspace{5pt}

\begin{tablenotes}[para]
\footnotesize
\begin{minipage}{\linewidth}
\justifying
ML denotes monolayer. FE and AFE denote ferroelectric and antiferroelectric ordering, respectively.
DP refers to deformation-potential mechanism. IPE and PVE denote the inverse piezoelectric effect
and photovoltaic effect, respectively. Photostriction in BiFeO$_3$ was measured along pseudocubic
[100] directions, not along the polar [111] axis
\cite{SM_zhang2025giant,SM_wen2013electronic,SM_kundys2010light}.
\end{minipage}
\end{tablenotes}
\end{threeparttable}

\end{table*}

\clearpage

\section{Growth}
SnS films were grown via physical vapour deposition (PVD) at atmospheric pressure using a home-built single-zone furnace equipped with two mass flow controllers. SnS powder (99.5\% purity, Thermo Scientific), placed in an alumina boat inside a 1-inch quartz tube at the furnace center, served as the precursor. Freshly cleaved mica or $5 \times 5$~mm$^2$ diced MgO(110) substrates were placed downstream. To prevent unintended growth during the temperature ramp, a reverse flow of Ar (from substrate toward powder) was maintained. Upon reaching the growth temperature of $800\,^{\circ}\mathrm{C}$, the gas flow was switched to the downstream direction at a rate of 80~sccm for 10~s. The growth was subsequently quenched, followed by a 100~sccm reverse flow of Ar.

\section{Atomic Force Microscopy}
The thickness of the as-grown samples was characterized using a Bruker Dimension Icon atomic force microscope (AFM) in tapping mode.

\section{Polarized Optical Microscopy}
Bright-field cross-polarized microscopy was performed in reflection geometry using an Olympus BX60 microscope equipped with a $50\times$ objective lens (Olympus MSPlan 50, NA = 0.8). The sample was illuminated with a broadband white light source and imaged using a CCD camera (ACCU-SCOPE, AU-600-HD). A polarizer (Olympus U-PO) and an analyzer (Olympus U-AN360) were placed in the optical path of the incident light and the camera, respectively, in a cross-polarized configuration. The samples were mounted on a rotating stage to acquire angular-dependent polarized optical images.

\section{Second-Harmonic Generation}
Fundamental pulsed laser (Tsunami, Spectra-Physics) with a center wavelength of 800~nm and a pulse width of $<100$~fs was used to collect second-harmonic generation (SHG) signal in a reflection geometry. Details of the experimental setup and SHG angular anisotropy measurements are reported in Ref.~\cite{SM_puri2024substrate}. For SHG spatial mapping, the sample position was controlled using motorized XY translation stages (Zaber Technologies).

\section{Pump-Probe Spectroscopy}
Schematics of the pump-probe setup is illustrated in Fig.~\ref{fig:S7}. A Ti:Sapphire laser (Tsunami, Spectra-Physics) generating pulses of $<100$~fs duration and 80~MHz repetition rate was used. The fundamental output at 800~nm was split into a pump and a probe beam with an intensity ratio of 2:1, which produced a fluence at the sample of 0.29 mJ/cm$^2$ and 0.15 mJ/cm$^2$, respectively. The pump beam path was fixed using a static retroreflector (Edmund Optics, \#15-901), while the probe beam passed through a half-wave plate (Thorlabs, WPH05M-808) for polarization control and reflected off a rapid-scanning optical delay line (A.P.E. scanDelay) to modulate the path length. The pump and probe beams were focused onto a $\sim$3~\textmu m spot on the sample using an objective lens (Olympus LUCPlanFLN 20$\times$, NA = 0.45). The reflected signal passed through a linear polarizer (Thorlabs, LPVISE100-A) to reject pump scatter and a long-pass filter to block SHG signals, before detection by a balanced photodiode (Thorlabs, PDB230A). The transient reflectivity signal was recorded using lock-in detection, with the output monitored via an oscilloscope for real-time decay dynamics. The spatial mapping was recorded by using the motorized XY stages (Zaber Technologies) translating the sample position.

\section{Static Reflectance Measurement}
Static reflectance was measured using the probe line of the pump-probe setup with the pump beam blocked. A horizontally polarized 800~nm probe beam (0.5~mW average power) was used. The sample was rotated to align the crystallographic armchair (AC) or zigzag (ZZ) axes with the laser polarization to measure the anisotropic reflectance. A plane mirror (Thorlabs, PF05-03-P01) with a known reflectivity of 0.965 was used as a reference standard.

\section{Density Functional Photostriction Calculation}
We performed Density Functional Theory (DFT) calculations with the plane wave open-source code \textsc{Abinit} (version 9.6)~\cite{SM_gonze2009abinit}. Bulk antiferroelectric SnS was modeled within the PBE approximation~\cite{perdew1996generalized} of the exchange-correlation functional. The Projector Augmented Wave (PAW)~\cite{SM_blochl1994projector} formalism was employed using projectors from the Jollet, Torrent, Holzwarth database (version 1.1)~\cite{SM_jollet2014generation}. We employ a plane wave cut-off of 18~Ha and a sampling of the Brillouin zone of $24 \times 24 \times 9$. Electronic convergence was reached when the residual of the Kohn-Sham potential reach values smaller than $10^{-14}$, meanwhile structural relaxation was performed until forces on the ions were smaller than $5\times 10^{-6}$~Ha/Bohr.

We modeled photostriction in bulk AFE SnS using a constrained occupation scheme approach~\cite{SM_paillard2019photoinduced,SM_verstraete2025abinit}. This method simulates the presence of thermalized photo-excited carriers within the electronic structure and allows the lattice to relax in response to the modified electronic configuration. The carrier concentration, denoted as $n_{c}$, was varied from 0 to 0.16 electrons per formula unit (e/u.c.). Note that the unit cell contains 8 atoms (4 Sn and 4 S). 

Specifically, conduction band states are constrained to host $n_{c}$ electrons and valence bands are constrained to host $n_{c}$ holes by introducing two separate quasi-Fermi levels $\mu_{e}$ and $\mu_{h}$ when populating the Kohn-Sham states during the electronic self-consistent field iterations. We practically find $\mu_{e}$ and $\mu_{h}$ by solving Eqs. (1) $\&$ (2) of Ref~\cite{SM_paillard2019photoinduced,SM_verstraete2025abinit} using a bisection algorithm, adopting a Fermi-Dirac smearing scheme with an electronic temperature of 0.003 Ha.

For computing efficiency purposes, we employed the Vienna Ab initio Simulation Package (VASP) ~\cite{SM_kresse1993ab, SM_kresse1996efficient} to calculate the complex electronic dielectric constant in the Independent Particle Approximation, with a 500~eV plane wave cut-off, a $24 \times 24 \times 9$ mesh of the Brillouin zone and the SCAN exchange-correlation functional (which provides an improved electronic bandgap compared to PBE). The SCF cycle was converged when energy differences between two iterations were smaller than $10^{-7}$~eV. The structure was geometrically (ionic position, cell shape and volume) relaxed until forces on the ions were smaller than $10^{-3}$~eV/\AA. The orthorombic relaxed cell was then deformed sequentially by applying a multiplicative coefficient $(1+\eta)$ to each axis (the AC direction, the ZZ direction and the out-of-plane direction), the ionic position were relaxed and the real and imaginary part of the relative dielectric permittivity were calculated. The obtained dielectric permittivity versus uniaxial strain is depicted in Fig.~\ref{fig:S6}.

\section{Thin-film interference modelling for reflectance and transient reflectivity}
Let us consider a simple three-layer system (as shown in Fig. 5a of the main text): air of refractive index $n_0$ = 1, SnS film of refractive index $n+ik$, and thickness $d$  on the MgO substrate with refractive index $n_s$ = 1.7276 for 800 nm ~\cite{SM_0polyanskiy2024refractiveindex}. We consider a monochromatic light of vacuum wavelength $\lambda$ is incident normally on the SnS film. The incident light is partially reflected from the air-SnS interface, and partially transmitted. Inside the SnS film, the light travels with the wavenumber:
\begin{equation}
    k_{SnS} = \frac{2\pi(n+ik)}{\lambda}
\end{equation}

The total phase accumulated during the single pass through the SnS film of thickness \textit{d} is given by:
\begin{equation}
    \phi = \frac{2\pi(n+ik)}{\lambda} \cdot d
\end{equation}

Accounting for multiple internal reflections within the SnS film, the light reflected from the SnS-MgO interface returns to SnS-air interface,  acquiring the phase factor of $e^{2i\phi }$ per round trip.

At normal incidence,  the Fresnel reflection and transmission coefficients for the air-SnS interface is~\cite{SM_0hecht2012optics}:
\begin{equation}
    r_{01} = \frac{n_0 - (n+ik)}{n_0 + (n+ik)}
\end{equation}

\begin{equation}
    t_{01} = \frac{2n_0}{n_0 + (n+ik)}
\end{equation}

Similarly,  the reflection coefficient for the SnS-MgO interface is:
\begin{equation}
    r_{12} = \frac{(n+ik) - n_s}{(n+ik) + n_s}
\end{equation}

If $E_0$ be the incident field amplitude, the total reflected amplitude is:
\begin{equation}
    E_r = r_{01} \cdot E_0+\frac{t_{01} \cdot r_{12} \cdot t_{10} \cdot e^{2i\phi }}{1-r_{10} \cdot r_{12} \cdot e^{2i\phi }} \cdot E_0
\end{equation}

Since $r_{01}=-r_{10}$ and $t_{01} \cdot t_{10}=1-r_{01}^{2}$,
the net reflection coefficient for the air-SnS-MgO structure simplifies to:
\begin{equation}
    r_{net}(\lambda,d) = \frac{r_{01}+r_{12} \cdot e^{2i\phi }}{1+r_{01} \cdot r_{12} \cdot e^{2i\phi }}
\end{equation}

The total reflectance for the air-SnS-MgO structure is therefore:
\begin{equation}
    R(\lambda,d) = \left|{\frac{r_{01}+r_{12} \cdot e^{2i\phi }}{1+r_{01} \cdot r_{12} \cdot e^{2i\phi }}}\right|^2
\end{equation}

This is the static reflection. \\
Upon pump excitation,  the complex refractive index of the SnS film is modulated by $\Delta n$ and $\Delta k$. The perturbed refractive index of the SnS film  is:
\begin{equation}
    n' = (n + \Delta n) + i(k + \Delta k)
\end{equation}

and the corresponding phase shift is:
\begin{equation}
    \phi' = \frac{2 \pi[(n + \Delta n) + i(k + \Delta k)]}{\lambda}.d
\end{equation}

The resultant reflectivity due to the perturbed refractive index is:
\begin{equation}
    R'(\lambda,d) = \left|{\frac{r'_{01}+r'_{12} \cdot e^{2i\phi'}}{1+r'_{01} \cdot r'_{12} \cdot e^{2i\phi'}}}\right|^2
\end{equation}

where, $r'_{01} = \frac{n_0 - n'}{n_0 + n'}$ and $r'_{12} = \frac{n' - n_s}{n' + n_s}$

The transient differential signal measured in the pump-probe spectroscopy is:
\begin{equation}
    \frac{\Delta R}{R} = \frac{R'(\lambda,d)-R(\lambda,d)}{R(\lambda,d)}
\end{equation}
Given that $\Delta n << n$  and $\Delta k << k$,   the perturbed reflectance $R'$ can be approximated using a first-order multivariate Taylor expansion around the static reflectance $R(\lambda, d)$ as:
\begin{equation}
R' \approx R + \left( \frac{\partial R}{\partial n} \right) \cdot \Delta n + \left( \frac{\partial R}{\partial k} \right) \cdot \Delta k
\end{equation}
Consequently,  the transient reflectivity change simplifies to:
\begin{equation}
\frac{\Delta R}{R} = \frac{R'-R}{R} = \frac{1}{R} \cdot  \left(\frac{\partial R}{\partial n} \right) \cdot \Delta n + \frac{1}{R} \left( \frac{\partial R}{\partial k} \right) \cdot \Delta k
\end{equation}
On differentiating $R$ with respect to $n$ and $k$ and substituting the expressions back in Eq.  S14 yields:
\begin{equation}
\begin{split}
\frac{\Delta R}{R} = & \frac{2}{R} Re \left[ r_{\text{net}}^{*} \cdot \frac{\frac{-2n_{0}}{\left(n_0+(n+ik)\right)^{2}}+\left(\frac{2n_s}{\left((n+ik)+n_s\right)^{2}}+i \cdot \frac{4\pi d}{\lambda} \cdot r_{12}\right) \cdot (1 - r_{01}^{2}) \cdot e^{2i\phi} +\frac{2n_{0}}{\left(n_0+(n+ik)\right)^{2}} \cdot r_{12}^{2} \cdot e^{4i\phi}}{{\left(1+r_{01} \cdot r_{12} \cdot e^{2i\phi} \right)}^{2}} \right] \cdot \Delta n \\
& + \frac{2}{R} Re \left[ r_{\text{net}}^{*} \cdot i \frac{\frac{-2n_{0}}{\left(n_0+(n+ik)\right)^{2}}+\left(\frac{2n_s}{\left((n+ik)+n_s\right)^{2}}+i \cdot \frac{4\pi d}{\lambda} \cdot r_{12}\right) \cdot (1 - r_{01}^{2}) \cdot e^{2i\phi} +\frac{2n_{0}}{\left(n_0+(n+ik)\right)^{2}} \cdot r_{12}^{2} \cdot e^{4i\phi}}{{\left(1+r_{01} \cdot r_{12} \cdot e^{2i\phi} \right)}^{2}} \right] \cdot \Delta k
\end{split}
\end{equation}
Considering only the contribution from the real refractive index modulation ($\Delta n$), the expression reduces to:
\begin{equation}
\frac{\Delta R}{R} =  \frac{2}{R} Re \left[r_{net}^{*} \cdot \frac{\frac{-2n_{0}}{\left(n_0+(n+ik)\right)^{2}}+\left(\frac{2n_s}{\left((n+ik)+n_s\right)^{2}}+i \cdot \frac{4\pi d}{\lambda} \cdot r_{12}\right) \cdot (1 - r_{01}^{2}) \cdot e^{2i\phi} +\frac{2n_{0}}{\left(n_0+(n+ik)\right)^{2}} \cdot r_{12}^{2} \cdot e^{4i\phi}}{{\left(1+r_{01} \cdot r_{12} \cdot e^{2i\phi} \right)}^{2}} \right] \cdot \Delta n
\end{equation}
This expression demonstrates that the measured differential reflectance scales linearly with the real refractive index change ($\Delta n$), modulated by a complex function that accounts for multiple internal reflections within the thin-film cavity.

\newpage
\section{Estimation of photo-induced temperature rise ($\Delta$T) and thermal strain ($\epsilon_{\lowercase{th}}$)}
To qualitatively evaluate the potential thermal contribution to the transient reflectivity response, we estimate the pump-induced local temperature rise $\Delta$T in the SnS sample using the relation:
\begin{equation}
    \Delta T = \frac{F \cdot  \alpha \cdot (1-R)}{\rho \cdot C_p},
\end{equation}
where  $F$ is pump fluence, $\alpha$ is absorption coefficient,  $R$ is reflectivity, $\rho$ is mass density,  and $C_{p}$ is specific heat capacity. For our calculations, we use the known physical constants for SnS: $\rho$ = 5.22 $\text{gm/cm}^{3}$ and $C_{p}$ = 250 mJ/(gm.K).\\
Using the experimental pump fluence 0.29 $\text{mJ/cm}^{2}$ and the direction-dependent complex refractive indices ($n_{AC} = 4.23 + 0.12i$ and $n_{ZZ} = 4.36 + 0.25i$),  we estimate a maximum rise in temperature of 2.5 K and 5.1 K  along the AC and ZZ directions, respectively. The resulting thermal strain ($\epsilon_{th}$) is estimated using $\epsilon_{th} (strain) = \beta \Delta T$,  where $\beta$ is the thermal expansion coefficient ($2.8 \times 10^{-7}\text{K}^{-1}$ for SnS) ~\cite{SM_madelung2004semiconductors}.  This yields a thermal strain of $\sim 10^{-7}$ along the AC direction and $\sim 10^{-6}$ along the ZZ direction. \\
By utilizing the elasto-optic coefficients extracted from Fig.~\ref{fig:S6},  we converted the extracted refractive index modulations ($\Delta n_{AC} = -0.00302$ and $\Delta n_{ZZ} = +0.00022$) into their corresponding strain values of $+3.2 \times 10^{-4}$ and $-1.9 \times 10^{-4}$ for the AC and ZZ directions, respectively.  This conversion reveals that the thermal strains are at least two orders of magnitude smaller than the required strain to produce the observed modulation of refractive indices.  Furthermore, thermal expansion is inherently isotropic~\cite{SM_luo2023ultrafast} and positive-valued; it cannot account for the positive and negative strain observed along the distinct crystallographic axes of SnS. 

\section{Estimation of strain corresponding to the extracted $\Delta \lowercase{n}$}
The extracted $\Delta n$ along the AC and ZZ directions are $-$0.00302 and $+$0.00022, respectively.  To determine the corresponding lattice strain,  we first calculate the total change in dielectric permittivity ($\Delta \epsilon$).  The relation between the change in dielectric permittivity and the change in refractive index is obtained by differentiating the standard optical relation: $\epsilon = n^{2}$,  yielding $\Delta \epsilon \approx 2n \times \Delta n$.  For AC direction,  the dielectric permittivity change corresponding to $\Delta n_{AC}$ = $-$0.00302 is:
\begin{equation}
\Delta \epsilon_{AC} = 2n_{AC} \times \Delta n_{AC} = 2 \times (4.23) \times (-0.00302) = -0.0255
\end{equation}
Now, strain corresponding to this change in dielectric permittivity can be obtained using the slope of $\epsilon'_{yy}$ vs. $\Delta b/b (\%)$ plot of Fig. \ref{fig:S6} using relation:
\begin{equation}
Strain_{AC} = \frac{\Delta \epsilon_{AC}}{Slope_{AC}} =\frac{-0.0255}{-80} =+ 0.00032 = +0.032\%
\end{equation}
Similarly,  for ZZ direction,  the dielectic permittivity change corresponding to $\Delta n_{ZZ}$ = +0.00022 is:
\begin{equation}
\Delta \epsilon_{ZZ} = 2n_{ZZ} \times \Delta n_{ZZ} = 2 \times (4.36) \times (+0.00022) = 0.0019
\end{equation}
Now, strain corresponding to this change in dielectric permittivity can be obtained using the slope of $\epsilon'_{xx}$ vs.  $\Delta a/a (\%)$ plot of Fig. \ref{fig:S6} using relation:
\begin{equation}
Strain_{ZZ} = \frac{\Delta \epsilon_{ZZ}}{Slope_{ZZ}} =\frac{0.00191}{-10} = -0.00019 = - 0.019\%
\end{equation}

 \section{Estimation of $\Delta \lowercase{n}$ from DFT-calculated photostriction}
 
 Here,  we estimate $\Delta n$ expected from a representative carrier density of 0.05 $e\text{-}h/\text{u.c.}$ explicitly used in the DFT calculation [Fig. 5(a)].  Note that this corresponds to a high carrier concentration of $n_{c}$ = 2.43 $\times$ $\text{10}^{20}$ $\text{cm}^{-3}$.  At this carrier density,  the induced lattice strains along the AC and ZZ directions are $\sim +0.15\%$ and $\sim -0.6\%$, respectively. Along the AC direction, we utilize the slope of $\epsilon'_{yy}$ vs.  $\Delta b/b (\%)$ from Fig.~\ref{fig:S6} and $\Delta n_{AC}$ is estimated using the relation:
\begin{equation}
\Delta n_{AC} = \frac{\Delta \epsilon_{AC}}{2 n_{AC}} = \frac{\text{slope}_{AC} \times \text{strain}_{AC}}{2 n_{AC}} = \frac{(-80) \times (0.0015)}{2 \times 4.23} = -0.0142
\end{equation}
Similarly, along the ZZ direction, we employ the slope of the $\epsilon'_{xx}$ vs. $\Delta a/a(\%)$ plot from Fig.~\ref{fig:S6} to estimate $\Delta n_{ZZ}$ as follows:
\begin{equation}
\Delta n_{ZZ} = \frac{\Delta \epsilon_{ZZ}}{2 n_{ZZ}} = \frac{\text{slope}_{ZZ} \times \text{strain}_{ZZ}}{2 n_{ZZ}} = \frac{(-10) \times (-0.006)}{2 \times 4.36} = +0.0069
\end{equation}

\section{Estimation of Deformation potential}
The photoinduced strain ($\epsilon$) in a semiconductor is governed by the balance between the electronic pressure generated by population of photo-excited carriers $n_{c}$ and the restorative lattice stiffness.  So,  the deformation potential coupling constant $D_{DP}$ can be expressed as ~\cite{SM_ruello2015physical}:
\begin{equation}
D_{\text{DP}} = \frac{\epsilon (strain) \cdot C_{ii}}{n_{\text{c}}}
\end{equation}
where, $C_{ii}$ is the elastic constant of the material. 
We estimate $n_{c}$ using the relation:  
\begin{equation}
n_{c} = \frac{F \cdot \alpha \cdot (1-R)}{E_{photon}}
\end{equation}
Here,  $F$ is pump fluence, $\alpha$ is absorption coefficient,  $R$ is reflectivity,  and $E_{photon}$ is pump photon energy.  Using our experimental parameters:  $F = 0.29~\text{mJ/cm}^2$, $\alpha = 1.88 \times 10^4 ~\text{cm}^{-1}$  (for AC) and $3.85 \times 10^{4} ~\text{cm}^{-1}$ (for ZZ), $R = 0.3817$ (for AC) and $0.39$ (for ZZ), $E_{photon}$ = 1.55 eV,  the photoexcited carrier concentration for AC and ZZ directions are 1.36 $\times$ $\text{10}^{19}$ $\text{cm}^{-3}$ and 2.75 $\times$ $\text{10}^{19}$ $\text{cm}^{-3}$, respectively.  Using elastic stiffness constant: $C_{11}$ = 33 GPa (for ZZ),  and $C_{22}$ = 38 GPa (for AC)~\cite{SM_osti_1197543}, for the $0.032\%$ of strain along the AC direction, the required $D_{DP,AC}$ is estimated to be 5.58 eV.  Similarly,  for $0.019\%$ of strain along the ZZ direction,  $D_{DP,ZZ}$ is estimated to be 1.43 eV.  These values are within the physically expected regime for group-IV monochalcogenides,  providing the robust validation of the mechanism.  Moreover,  our experimental low-density strain scales linearly and interpolates perfectly with the 0.5\% strain predicted by the high-denisty DFT calculation (0.05 $e\text{-}h/\text{u.c.}$), confirming an intrinsic structural-electronic coupling.

\section{Estimation of Internal screening field in Inverse piezoelectric effect}

In the inverse piezoelectric effect (IPE) mechanism, the induced lattice strain ($\epsilon_{ij}$) is linearly proportional to the internal electric field ($E_{k}$) via the piezoelectric tensor ($d_{kij}$) as ~\cite{SM_nye1985physical}.:
\begin{equation}
\epsilon_{ij} (strain) = d_{kij} \cdot  E_{k}
\end{equation}
For AC direction,  using the strain $\epsilon_{AC} = 0.032\%$ and the piezoelectric coefficient $d_{222} = 22.07$ pm/V ~\cite{SM_fei2015giant}, we estimate the internal screening field along the AC direction to be to be:
\begin{equation}
E_{AC} = \frac{\epsilon_{AC}}{d_{222}} = \frac{3.2 \times 10^{-4}}{22.07 \times 10^{-12} \text{ m/V}} \approx 14.5 \text{ MV/m}
\end{equation}

To show that this magnitude of shielding field is physically unrealistic for our multilayer samples in our pump-probe measurement, we calculate the electric shielding field generated from photo-excited carriers.
The steady-state photo-Dember voltage ($V_D$) is described by the following equation~\cite{SM_gu2002study}:
\begin{equation}
V_D = \frac{k_B T}{e} \frac{b-1}{b+1} \ln \left( 1 + \frac{(b+1)\Delta n}{n_0 b + p_0} \right),
\end{equation}
where $k_B$ is the Boltzmann constant, $T$ is the temperature of the photoexcited electrons and holes, and $e$ is the elementary charge. The parameter $b = \mu_e / \mu_p$ is the mobility ratio of the electrons ($\mu_e$) and holes ($\mu_p$). The terms $n_0$ and $p_0$ represent the initial density of the electrons and holes, respectively, while $\Delta n$ is the photoinjected excess carrier density.

In the high-injection state (saturation limit), the generated electric field ($E$) can be approximated as:
\begin{equation}
E \approx \frac{k_B T}{e} \alpha,
\end{equation}
where $\alpha$ is the absorption coefficient. Using $\alpha = 1.88 \times 10^{-6}$ (for the AC direction of our SnS), the maximum generated field is $0.0486$ MV/m. Therefore, a $15$ MV/m field is unrealistic to be generated solely from photo-injection.

\clearpage

% ---------------------------------------------------------
% SUPPLEMENTARY FIGURES
% ---------------------------------------------------------

% Fig S1: single power dependence -> 0.5
\begin{figure}[htbp]
    \centering
    \includegraphics[width=0.5\linewidth]{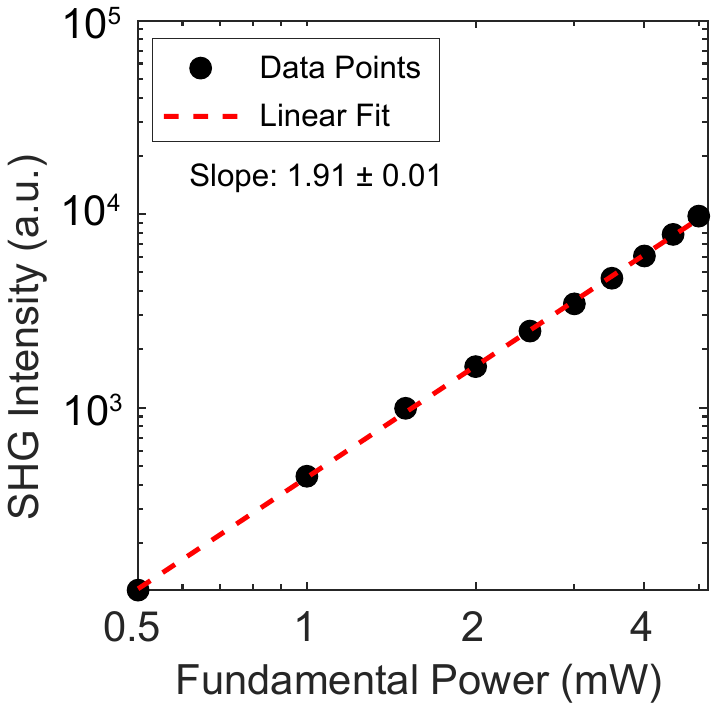}
    \caption{Power dependent SHG measurement. The figure shows a double logarithmic plot of SHG dependence on pump power. The black dots represent the experimental data points, and the red dashed line represents the linear fit. The slope of the linear fit is 1.91 $\pm$ 0.01($\approx$ 2), which confirms the quadratic dependence of SHG on pump power.}
    \label{fig:S1}
\end{figure}

% Fig S2: Multi-panel -> 0.8
\begin{figure}[h]
    \centering
    \includegraphics[width=0.8\linewidth]{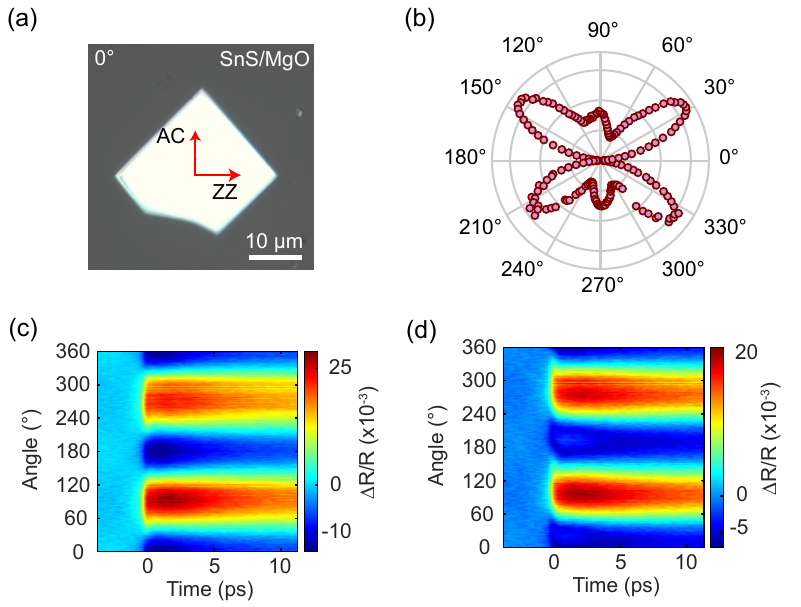}
    \caption{Pump polarization dependence of ultrafast reflectivity. (a) Optical image of monodomain 275 nm thick SnS island showing the AC and ZZ direction. (b) SHG polar plot taken in parallel configuration for the 275 nm thick SnS island.  (c,d) Polarization-resolved transient reflectivity change ($\Delta$R/R) of 275 nm thick SnS island for (c) parallel (pump$\parallel$probe) configuration and (d) perpendicular (pump$\perp$probe) configurations. The observed sign of $\Delta$R/R remains identical in each case, demonstrating that pump polarization direction does not influence the sign. Also, this 275 nm thick SnS shows $\Delta$R/R $>$ 0 for probe$\parallel$AC (Angle=90°) and $\Delta$R/R $<$ 0 for probe$\parallel$ZZ (Angle = 0°), exact opposite in sign compared to the data from 49.5 nm thick sample shown in Fig.4b of the main text. This originates from optical interference, as described in the main text and Supplementary Text.}
    \label{fig:S2}
\end{figure}

% Fig S3: Multi-panel -> 0.8
\begin{figure}[h]
    \centering
    \includegraphics[width=0.8\linewidth]{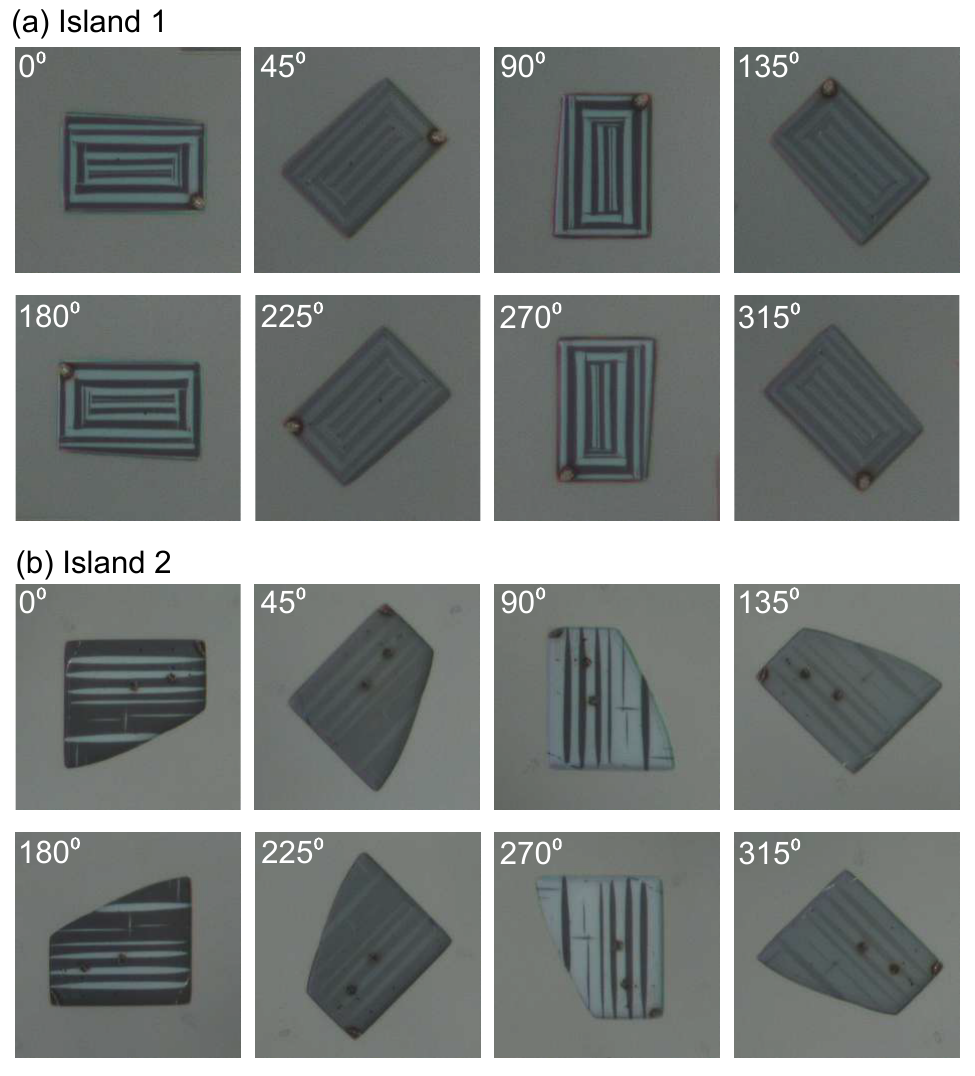}
    \caption{Cross-polarized microscopy images. Cross-polarization microscopy images of multidomain SnS at 45° step angle for two different islands: (a) Island 1 and (b) Island 2.}
    \label{fig:S3}
\end{figure}

% Fig S4: Multi-panel -> 0.8
\begin{figure}[h]
    \centering
    \includegraphics[width=0.8\linewidth]{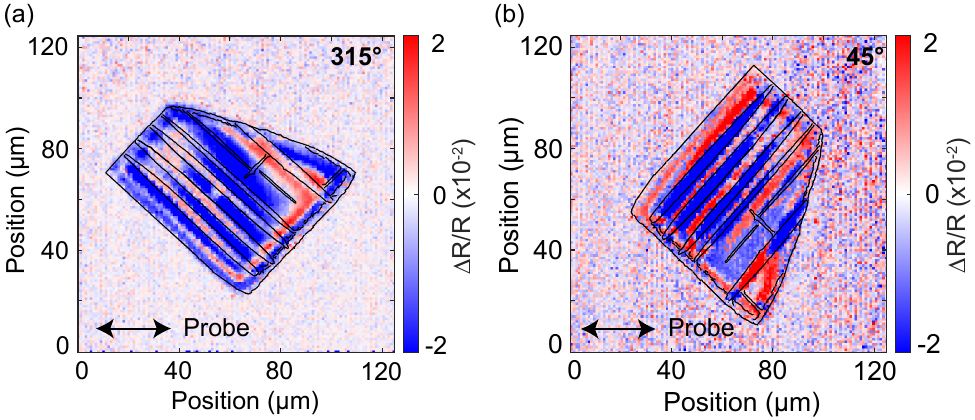}
    \caption{Pump-probe signal mapping. Ultrafast transient reflection signal mapping on multidomain SnS island 2 (Fig. \ref{fig:S3}b) for two different orientations: (a) 315° and (b) 45°.}
    \label{fig:S4}
\end{figure}

\clearpage

% Fig S5: Multi-panel -> 0.8
\begin{figure}[h]
    \centering
    \includegraphics[width=0.8\linewidth]{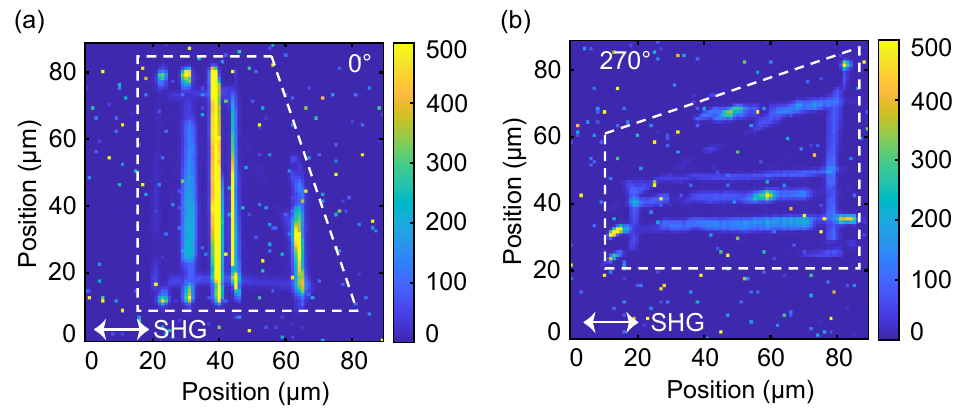}
    \caption{SHG mapping on multidomain SnS island 2 (Fig. \ref{fig:S3}b) for two different orientations: (a) 0° and (b) 270°.}
    \label{fig:S5}
\end{figure}

\clearpage

% Fig S6: Pump probe setup
\begin{figure}[h]
    \centering
    \includegraphics[width=0.8\linewidth]{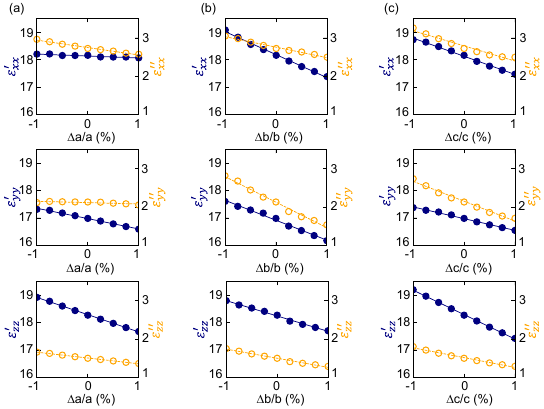}
    \caption{Elasto-optic effect in SnS.  Change in relative dielectric permittivity, calculated by DFT in the Independent Particle Approximation, at a photon energy of $\hbar \omega$  = 1.55 eV as a function of imposed uniaxial strain. The plots are arranged in three columns corresponding to the direction of the applied strain: (a) along the ZZ direction (first column), (b) along the AC direction (second column), and (c) along the out-of-plane direction (third column). For each case, the strain is varied independently from -1$\%$ to +1$\%$.}
    \label{fig:S6}
\end{figure}

\clearpage

% Fig S7: Pump probe setup
\begin{figure}[h]
    \centering
    \includegraphics[width=0.8\linewidth]{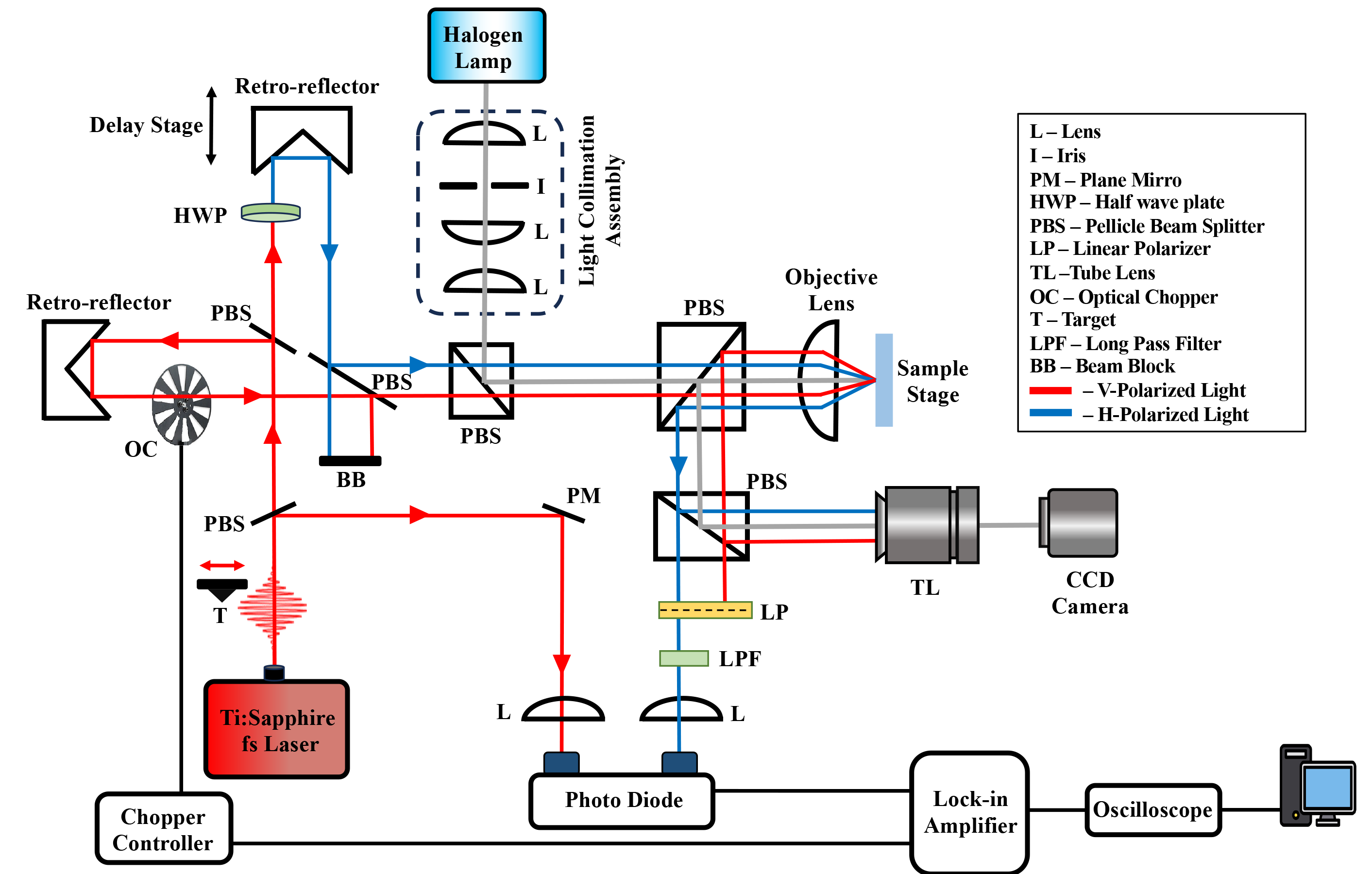}
    \caption{Schematic diagram of degenerate pump-probe reflection spectroscopy.}
    \label{fig:S7}
\end{figure}

\clearpage

% Reset the bibliography counter for the SM
\makeatletter
\setcounter{NAT@ctr}{0}
\makeatother


\begin{thebibliography}{48}%
\makeatletter
\providecommand \@ifxundefined [1]{%
 \@ifx{#1\undefined}
}%
\providecommand \@ifnum [1]{%
 \ifnum #1\expandafter \@firstoftwo
 \else \expandafter \@secondoftwo
 \fi
}%
\providecommand \@ifx [1]{%
 \ifx #1\expandafter \@firstoftwo
 \else \expandafter \@secondoftwo
 \fi
}%
\providecommand \natexlab [1]{#1}%
\providecommand \enquote  [1]{``#1''}%
\providecommand \bibnamefont  [1]{#1}%
\providecommand \bibfnamefont [1]{#1}%
\providecommand \citenamefont [1]{#1}%
\providecommand \href@noop [0]{\@secondoftwo}%
\providecommand \href [0]{\begingroup \@sanitize@url \@href}%
\providecommand \@href[1]{\@@startlink{#1}\@@href}%
\providecommand \@@href[1]{\endgroup#1\@@endlink}%
\providecommand \@sanitize@url [0]{\catcode `\\12\catcode `\$12\catcode
  `\&12\catcode `\#12\catcode `\^12\catcode `\_12\catcode `\%12\relax}%
\providecommand \@@startlink[1]{}%
\providecommand \@@endlink[0]{}%
\providecommand \url  [0]{\begingroup\@sanitize@url \@url }%
\providecommand \@url [1]{\endgroup\@href {#1}{\urlprefix }}%
\providecommand \urlprefix  [0]{URL }%
\providecommand \Eprint [0]{\href }%
\providecommand \doibase [0]{https://doi.org/}%
\providecommand \selectlanguage [0]{\@gobble}%
\providecommand \bibinfo  [0]{\@secondoftwo}%
\providecommand \bibfield  [0]{\@secondoftwo}%
\providecommand \translation [1]{[#1]}%
\providecommand \BibitemOpen [0]{}%
\providecommand \bibitemStop [0]{}%
\providecommand \bibitemNoStop [0]{.\EOS\space}%
\providecommand \EOS [0]{\spacefactor3000\relax}%
\providecommand \BibitemShut  [1]{\csname bibitem#1\endcsname}%
\let\auto@bib@innerbib\@empty
%</preamble>
\bibitem [{\citenamefont {Wu}\ and\ \citenamefont
  {Zeng}(2016)}]{wu2016intrinsic}%
  \BibitemOpen
  \bibfield  {author} {\bibinfo {author} {\bibfnamefont {M.}~\bibnamefont
  {Wu}}\ and\ \bibinfo {author} {\bibfnamefont {X.~C.}\ \bibnamefont {Zeng}},\
  }\bibfield  {title} {\bibinfo {title} {Intrinsic ferroelasticity and/or
  multiferroicity in two-dimensional phosphorene and phosphorene analogues},\
  }\href {https://doi.org/10.1021/acs.nanolett.6b00726} {\bibfield  {journal}
  {\bibinfo  {journal} {Nano Lett.}\ }\textbf {\bibinfo {volume} {16}},\
  \bibinfo {pages} {3236} (\bibinfo {year} {2016})}\BibitemShut {NoStop}%
\bibitem [{\citenamefont {Wang}\ and\ \citenamefont
  {Qian}(2017{\natexlab{a}})}]{wang2017two}%
  \BibitemOpen
  \bibfield  {author} {\bibinfo {author} {\bibfnamefont {H.}~\bibnamefont
  {Wang}}\ and\ \bibinfo {author} {\bibfnamefont {X.}~\bibnamefont {Qian}},\
  }\bibfield  {title} {\bibinfo {title} {Two-dimensional multiferroics in
  monolayer group {IV} monochalcogenides},\ }\href
  {https://doi.org/10.1088/2053-1583/4/1/015042} {\bibfield  {journal}
  {\bibinfo  {journal} {2D Mater.}\ }\textbf {\bibinfo {volume} {4}},\ \bibinfo
  {pages} {015042} (\bibinfo {year} {2017}{\natexlab{a}})}\BibitemShut
  {NoStop}%
\bibitem [{\citenamefont {Bao}\ \emph {et~al.}(2019)\citenamefont {Bao},
  \citenamefont {Song}, \citenamefont {Liu}, \citenamefont {Chen},
  \citenamefont {Zhu}, \citenamefont {Abdelwahab}, \citenamefont {Su},
  \citenamefont {Fu}, \citenamefont {Chi}, \citenamefont {Yu} \emph
  {et~al.}}]{bao2019gate}%
  \BibitemOpen
  \bibfield  {author} {\bibinfo {author} {\bibfnamefont {Y.}~\bibnamefont
  {Bao}}, \bibinfo {author} {\bibfnamefont {P.}~\bibnamefont {Song}}, \bibinfo
  {author} {\bibfnamefont {Y.}~\bibnamefont {Liu}}, \bibinfo {author}
  {\bibfnamefont {Z.}~\bibnamefont {Chen}}, \bibinfo {author} {\bibfnamefont
  {M.}~\bibnamefont {Zhu}}, \bibinfo {author} {\bibfnamefont {I.}~\bibnamefont
  {Abdelwahab}}, \bibinfo {author} {\bibfnamefont {J.}~\bibnamefont {Su}},
  \bibinfo {author} {\bibfnamefont {W.}~\bibnamefont {Fu}}, \bibinfo {author}
  {\bibfnamefont {X.}~\bibnamefont {Chi}}, \bibinfo {author} {\bibfnamefont
  {W.}~\bibnamefont {Yu}}, \emph {et~al.},\ }\bibfield  {title} {\bibinfo
  {title} {Gate-tunable in-plane ferroelectricity in few-layer {SnS}},\ }\href
  {https://doi.org/10.1021/acs.nanolett.9b01419} {\bibfield  {journal}
  {\bibinfo  {journal} {Nano Lett.}\ }\textbf {\bibinfo {volume} {19}},\
  \bibinfo {pages} {5109} (\bibinfo {year} {2019})}\BibitemShut {NoStop}%
\bibitem [{\citenamefont {Higashitarumizu}\ \emph {et~al.}(2020)\citenamefont
  {Higashitarumizu}, \citenamefont {Kawamoto}, \citenamefont {Lee},
  \citenamefont {Lin}, \citenamefont {Chu}, \citenamefont {Yonemori},
  \citenamefont {Nishimura}, \citenamefont {Wakabayashi}, \citenamefont
  {Chang},\ and\ \citenamefont {Nagashio}}]{higashitarumizu2020purely}%
  \BibitemOpen
  \bibfield  {author} {\bibinfo {author} {\bibfnamefont {N.}~\bibnamefont
  {Higashitarumizu}}, \bibinfo {author} {\bibfnamefont {H.}~\bibnamefont
  {Kawamoto}}, \bibinfo {author} {\bibfnamefont {C.-J.}\ \bibnamefont {Lee}},
  \bibinfo {author} {\bibfnamefont {B.-H.}\ \bibnamefont {Lin}}, \bibinfo
  {author} {\bibfnamefont {F.-H.}\ \bibnamefont {Chu}}, \bibinfo {author}
  {\bibfnamefont {I.}~\bibnamefont {Yonemori}}, \bibinfo {author}
  {\bibfnamefont {T.}~\bibnamefont {Nishimura}}, \bibinfo {author}
  {\bibfnamefont {K.}~\bibnamefont {Wakabayashi}}, \bibinfo {author}
  {\bibfnamefont {W.-H.}\ \bibnamefont {Chang}},\ and\ \bibinfo {author}
  {\bibfnamefont {K.}~\bibnamefont {Nagashio}},\ }\bibfield  {title} {\bibinfo
  {title} {Purely in-plane ferroelectricity in monolayer {SnS} at room
  temperature},\ }\href
  {https://doi.org/https://doi.org/10.1038/s41467-020-16291-9} {\bibfield
  {journal} {\bibinfo  {journal} {Nat. Commun.}\ }\textbf {\bibinfo {volume}
  {11}},\ \bibinfo {pages} {2428} (\bibinfo {year} {2020})}\BibitemShut
  {NoStop}%
\bibitem [{\citenamefont {Zhu}\ \emph {et~al.}(2021)\citenamefont {Zhu},
  \citenamefont {Zhong}, \citenamefont {Guo}, \citenamefont {Wang},
  \citenamefont {Chen}, \citenamefont {Huang}, \citenamefont {He},
  \citenamefont {Su},\ and\ \citenamefont {Loh}}]{zhu2021efficient}%
  \BibitemOpen
  \bibfield  {author} {\bibinfo {author} {\bibfnamefont {M.}~\bibnamefont
  {Zhu}}, \bibinfo {author} {\bibfnamefont {M.}~\bibnamefont {Zhong}}, \bibinfo
  {author} {\bibfnamefont {X.}~\bibnamefont {Guo}}, \bibinfo {author}
  {\bibfnamefont {Y.}~\bibnamefont {Wang}}, \bibinfo {author} {\bibfnamefont
  {Z.}~\bibnamefont {Chen}}, \bibinfo {author} {\bibfnamefont {H.}~\bibnamefont
  {Huang}}, \bibinfo {author} {\bibfnamefont {J.}~\bibnamefont {He}}, \bibinfo
  {author} {\bibfnamefont {C.}~\bibnamefont {Su}},\ and\ \bibinfo {author}
  {\bibfnamefont {K.~P.}\ \bibnamefont {Loh}},\ }\bibfield  {title} {\bibinfo
  {title} {Efficient and anisotropic second harmonic generation in few-layer
  {SnS} film},\ }\href {https://doi.org/https://doi.org/10.1002/adom.202101200}
  {\bibfield  {journal} {\bibinfo  {journal} {Adv. Opt. Mater.}\ }\textbf
  {\bibinfo {volume} {9}},\ \bibinfo {pages} {2101200} (\bibinfo {year}
  {2021})}\BibitemShut {NoStop}%
\bibitem [{\citenamefont {Banai}\ \emph {et~al.}(2016)\citenamefont {Banai},
  \citenamefont {Horn},\ and\ \citenamefont {Brownson}}]{banai2016review}%
  \BibitemOpen
  \bibfield  {author} {\bibinfo {author} {\bibfnamefont {R.}~\bibnamefont
  {Banai}}, \bibinfo {author} {\bibfnamefont {M.}~\bibnamefont {Horn}},\ and\
  \bibinfo {author} {\bibfnamefont {J.}~\bibnamefont {Brownson}},\ }\bibfield
  {title} {\bibinfo {title} {A review of tin ({II}) monosulfide and its
  potential as a photovoltaic absorber},\ }\href
  {https://doi.org/https://doi.org/10.1016/j.solmat.2015.12.001} {\bibfield
  {journal} {\bibinfo  {journal} {Sol. Energy Mater. Sol. Cells}\ }\textbf
  {\bibinfo {volume} {150}},\ \bibinfo {pages} {112} (\bibinfo {year}
  {2016})}\BibitemShut {NoStop}%
\bibitem [{\citenamefont {Maragkakis}\ \emph {et~al.}(2022)\citenamefont
  {Maragkakis}, \citenamefont {Psilodimitrakopoulos}, \citenamefont
  {Mouchliadis}, \citenamefont {Sarkar}, \citenamefont {Lemonis}, \citenamefont
  {Kioseoglou},\ and\ \citenamefont {Stratakis}}]{maragkakis2022nonlinear}%
  \BibitemOpen
  \bibfield  {author} {\bibinfo {author} {\bibfnamefont {G.~M.}\ \bibnamefont
  {Maragkakis}}, \bibinfo {author} {\bibfnamefont {S.}~\bibnamefont
  {Psilodimitrakopoulos}}, \bibinfo {author} {\bibfnamefont {L.}~\bibnamefont
  {Mouchliadis}}, \bibinfo {author} {\bibfnamefont {A.~S.}\ \bibnamefont
  {Sarkar}}, \bibinfo {author} {\bibfnamefont {A.}~\bibnamefont {Lemonis}},
  \bibinfo {author} {\bibfnamefont {G.}~\bibnamefont {Kioseoglou}},\ and\
  \bibinfo {author} {\bibfnamefont {E.}~\bibnamefont {Stratakis}},\ }\bibfield
  {title} {\bibinfo {title} {Nonlinear optical imaging of in-plane anisotropy
  in two-dimensional {SnS}},\ }\href
  {https://doi.org/https://doi.org/10.1002/adom.202102776} {\bibfield
  {journal} {\bibinfo  {journal} {Adv. Opt. Mater.}\ }\textbf {\bibinfo
  {volume} {10}},\ \bibinfo {pages} {2102776} (\bibinfo {year}
  {2022})}\BibitemShut {NoStop}%
\bibitem [{\citenamefont {Sarkar}\ \emph {et~al.}(2023)\citenamefont {Sarkar},
  \citenamefont {Konidakis}, \citenamefont {Gagaoudakis}, \citenamefont
  {Maragkakis}, \citenamefont {Psilodimitrakopoulos}, \citenamefont
  {Katerinopoulou}, \citenamefont {Sygellou}, \citenamefont {Deligeorgis},
  \citenamefont {Binas}, \citenamefont {Oikonomou} \emph
  {et~al.}}]{sarkar2023liquid}%
  \BibitemOpen
  \bibfield  {author} {\bibinfo {author} {\bibfnamefont {A.~S.}\ \bibnamefont
  {Sarkar}}, \bibinfo {author} {\bibfnamefont {I.}~\bibnamefont {Konidakis}},
  \bibinfo {author} {\bibfnamefont {E.}~\bibnamefont {Gagaoudakis}}, \bibinfo
  {author} {\bibfnamefont {G.}~\bibnamefont {Maragkakis}}, \bibinfo {author}
  {\bibfnamefont {S.}~\bibnamefont {Psilodimitrakopoulos}}, \bibinfo {author}
  {\bibfnamefont {D.}~\bibnamefont {Katerinopoulou}}, \bibinfo {author}
  {\bibfnamefont {L.}~\bibnamefont {Sygellou}}, \bibinfo {author}
  {\bibfnamefont {G.}~\bibnamefont {Deligeorgis}}, \bibinfo {author}
  {\bibfnamefont {V.}~\bibnamefont {Binas}}, \bibinfo {author} {\bibfnamefont
  {I.~M.}\ \bibnamefont {Oikonomou}}, \emph {et~al.},\ }\bibfield  {title}
  {\bibinfo {title} {Liquid phase isolation of {SnS} monolayers with enhanced
  optoelectronic properties},\ }\href
  {https://doi.org/https://doi.org/10.1002/advs.202201842} {\bibfield
  {journal} {\bibinfo  {journal} {Adv. Sci.}\ }\textbf {\bibinfo {volume}
  {10}},\ \bibinfo {pages} {2201842} (\bibinfo {year} {2023})}\BibitemShut
  {NoStop}%
\bibitem [{\citenamefont {Xin}\ \emph {et~al.}(2016)\citenamefont {Xin},
  \citenamefont {Zheng}, \citenamefont {Su}, \citenamefont {Li}, \citenamefont
  {Zhang}, \citenamefont {Feng},\ and\ \citenamefont {Pan}}]{xin2016few}%
  \BibitemOpen
  \bibfield  {author} {\bibinfo {author} {\bibfnamefont {C.}~\bibnamefont
  {Xin}}, \bibinfo {author} {\bibfnamefont {J.}~\bibnamefont {Zheng}}, \bibinfo
  {author} {\bibfnamefont {Y.}~\bibnamefont {Su}}, \bibinfo {author}
  {\bibfnamefont {S.}~\bibnamefont {Li}}, \bibinfo {author} {\bibfnamefont
  {B.}~\bibnamefont {Zhang}}, \bibinfo {author} {\bibfnamefont
  {Y.}~\bibnamefont {Feng}},\ and\ \bibinfo {author} {\bibfnamefont
  {F.}~\bibnamefont {Pan}},\ }\bibfield  {title} {\bibinfo {title} {Few-layer
  tin sulfide: a new black-phosphorus-analogue {2D} material with a sizeable
  band gap, odd--even quantum confinement effect, and high carrier mobility},\
  }\href {https://doi.org/10.1021/acs.jpcc.6b06673} {\bibfield  {journal}
  {\bibinfo  {journal} {J. Phys. Chem. C}\ }\textbf {\bibinfo {volume} {120}},\
  \bibinfo {pages} {22663} (\bibinfo {year} {2016})}\BibitemShut {NoStop}%
\bibitem [{\citenamefont {Huang}\ \emph {et~al.}(2017)\citenamefont {Huang},
  \citenamefont {Chen}, \citenamefont {Wang}, \citenamefont {Peng},
  \citenamefont {Qian},\ and\ \citenamefont {Wang}}]{huang2017layer}%
  \BibitemOpen
  \bibfield  {author} {\bibinfo {author} {\bibfnamefont {Y.}~\bibnamefont
  {Huang}}, \bibinfo {author} {\bibfnamefont {X.}~\bibnamefont {Chen}},
  \bibinfo {author} {\bibfnamefont {C.}~\bibnamefont {Wang}}, \bibinfo {author}
  {\bibfnamefont {L.}~\bibnamefont {Peng}}, \bibinfo {author} {\bibfnamefont
  {Q.}~\bibnamefont {Qian}},\ and\ \bibinfo {author} {\bibfnamefont
  {S.}~\bibnamefont {Wang}},\ }\bibfield  {title} {\bibinfo {title}
  {Layer-dependent electronic properties of phosphorene-like materials and
  phosphorene-based van der waals heterostructures},\ }\href
  {https://doi.org/https://doi.org/10.1039/c7nr01952a} {\bibfield  {journal}
  {\bibinfo  {journal} {Nanoscale}\ }\textbf {\bibinfo {volume} {9}},\ \bibinfo
  {pages} {8616} (\bibinfo {year} {2017})}\BibitemShut {NoStop}%
\bibitem [{\citenamefont {Thomsen}\ \emph {et~al.}(1986)\citenamefont
  {Thomsen}, \citenamefont {Grahn}, \citenamefont {Maris},\ and\ \citenamefont
  {Tauc}}]{thomsen1986surface}%
  \BibitemOpen
  \bibfield  {author} {\bibinfo {author} {\bibfnamefont {C.}~\bibnamefont
  {Thomsen}}, \bibinfo {author} {\bibfnamefont {H.~T.}\ \bibnamefont {Grahn}},
  \bibinfo {author} {\bibfnamefont {H.~J.}\ \bibnamefont {Maris}},\ and\
  \bibinfo {author} {\bibfnamefont {J.}~\bibnamefont {Tauc}},\ }\bibfield
  {title} {\bibinfo {title} {Surface generation and detection of phonons by
  picosecond light pulses},\ }\href
  {https://doi.org/https://doi.org/10.1103/PhysRevB.34.4129} {\bibfield
  {journal} {\bibinfo  {journal} {Phys. Rev. B}\ }\textbf {\bibinfo {volume}
  {34}},\ \bibinfo {pages} {4129} (\bibinfo {year} {1986})}\BibitemShut
  {NoStop}%
\bibitem [{\citenamefont {Zeiger}\ \emph {et~al.}(1992)\citenamefont {Zeiger},
  \citenamefont {Vidal}, \citenamefont {Cheng}, \citenamefont {Ippen},
  \citenamefont {Dresselhaus},\ and\ \citenamefont
  {Dresselhaus}}]{zeiger1992theory}%
  \BibitemOpen
  \bibfield  {author} {\bibinfo {author} {\bibfnamefont {H.~J.}\ \bibnamefont
  {Zeiger}}, \bibinfo {author} {\bibfnamefont {J.}~\bibnamefont {Vidal}},
  \bibinfo {author} {\bibfnamefont {T.~K.}\ \bibnamefont {Cheng}}, \bibinfo
  {author} {\bibfnamefont {E.~P.}\ \bibnamefont {Ippen}}, \bibinfo {author}
  {\bibfnamefont {G.}~\bibnamefont {Dresselhaus}},\ and\ \bibinfo {author}
  {\bibfnamefont {M.~S.}\ \bibnamefont {Dresselhaus}},\ }\bibfield  {title}
  {\bibinfo {title} {Theory for displacive excitation of coherent phonons},\
  }\href {https://doi.org/https://doi.org/10.1103/PhysRevB.45.768} {\bibfield
  {journal} {\bibinfo  {journal} {Phys. Rev. B}\ }\textbf {\bibinfo {volume}
  {45}},\ \bibinfo {pages} {768} (\bibinfo {year} {1992})}\BibitemShut
  {NoStop}%
\bibitem [{\citenamefont {Daranciang}\ \emph {et~al.}(2012)\citenamefont
  {Daranciang}, \citenamefont {Highland}, \citenamefont {Wen}, \citenamefont
  {Young}, \citenamefont {Brandt}, \citenamefont {Hwang}, \citenamefont
  {Vattilana}, \citenamefont {Nicoul}, \citenamefont {Quirin}, \citenamefont
  {Goodfellow} \emph {et~al.}}]{daranciang2012ultrafast}%
  \BibitemOpen
  \bibfield  {author} {\bibinfo {author} {\bibfnamefont {D.}~\bibnamefont
  {Daranciang}}, \bibinfo {author} {\bibfnamefont {M.~J.}\ \bibnamefont
  {Highland}}, \bibinfo {author} {\bibfnamefont {H.}~\bibnamefont {Wen}},
  \bibinfo {author} {\bibfnamefont {S.~M.}\ \bibnamefont {Young}}, \bibinfo
  {author} {\bibfnamefont {N.~C.}\ \bibnamefont {Brandt}}, \bibinfo {author}
  {\bibfnamefont {H.~Y.}\ \bibnamefont {Hwang}}, \bibinfo {author}
  {\bibfnamefont {M.}~\bibnamefont {Vattilana}}, \bibinfo {author}
  {\bibfnamefont {M.}~\bibnamefont {Nicoul}}, \bibinfo {author} {\bibfnamefont
  {F.}~\bibnamefont {Quirin}}, \bibinfo {author} {\bibfnamefont
  {J.}~\bibnamefont {Goodfellow}}, \emph {et~al.},\ }\bibfield  {title}
  {\bibinfo {title} {Ultrafast photovoltaic response in ferroelectric
  nanolayers},\ }\href
  {https://doi.org/https://doi.org/10.1103/PhysRevLett.108.087601} {\bibfield
  {journal} {\bibinfo  {journal} {Phys. Rev. Lett.}\ }\textbf {\bibinfo
  {volume} {108}},\ \bibinfo {pages} {087601} (\bibinfo {year}
  {2012})}\BibitemShut {NoStop}%
\bibitem [{\citenamefont {Haleoot}\ \emph {et~al.}(2017)\citenamefont
  {Haleoot}, \citenamefont {Paillard}, \citenamefont {Kaloni}, \citenamefont
  {Mehboudi}, \citenamefont {Xu}, \citenamefont {Bellaiche},\ and\
  \citenamefont {Barraza-Lopez}}]{haleoot2017photostrictive}%
  \BibitemOpen
  \bibfield  {author} {\bibinfo {author} {\bibfnamefont {R.}~\bibnamefont
  {Haleoot}}, \bibinfo {author} {\bibfnamefont {C.}~\bibnamefont {Paillard}},
  \bibinfo {author} {\bibfnamefont {T.~P.}\ \bibnamefont {Kaloni}}, \bibinfo
  {author} {\bibfnamefont {M.}~\bibnamefont {Mehboudi}}, \bibinfo {author}
  {\bibfnamefont {B.}~\bibnamefont {Xu}}, \bibinfo {author} {\bibfnamefont
  {L.}~\bibnamefont {Bellaiche}},\ and\ \bibinfo {author} {\bibfnamefont
  {S.}~\bibnamefont {Barraza-Lopez}},\ }\bibfield  {title} {\bibinfo {title}
  {Photostrictive two-dimensional materials in the monochalcogenide family},\
  }\href {https://doi.org/https://doi.org/10.1103/PhysRevLett.118.227401}
  {\bibfield  {journal} {\bibinfo  {journal} {Phys. Rev. Lett.}\ }\textbf
  {\bibinfo {volume} {118}},\ \bibinfo {pages} {227401} (\bibinfo {year}
  {2017})}\BibitemShut {NoStop}%
\bibitem [{\citenamefont {Chen}\ \emph {et~al.}(2025)\citenamefont {Chen},
  \citenamefont {Liu}, \citenamefont {Guo}, \citenamefont {Yi} \emph
  {et~al.}}]{chen2025constructing}%
  \BibitemOpen
  \bibfield  {author} {\bibinfo {author} {\bibfnamefont {C.}~\bibnamefont
  {Chen}}, \bibinfo {author} {\bibfnamefont {W.}~\bibnamefont {Liu}}, \bibinfo
  {author} {\bibfnamefont {F.}~\bibnamefont {Guo}}, \bibinfo {author}
  {\bibfnamefont {Z.}~\bibnamefont {Yi}}, \emph {et~al.},\ }\bibfield  {title}
  {\bibinfo {title} {Constructing polymorphic phase boundary for
  high-performance inorganic photostrictive materials},\ }\href
  {https://doi.org/https://doi.org/10.1038/s41467-025-58100-1} {\bibfield
  {journal} {\bibinfo  {journal} {Nat. Commun.}\ }\textbf {\bibinfo {volume}
  {16}},\ \bibinfo {pages} {2788} (\bibinfo {year} {2025})}\BibitemShut
  {NoStop}%
\bibitem [{\citenamefont {Luo}\ \emph {et~al.}(2023)\citenamefont {Luo},
  \citenamefont {Zhang}, \citenamefont {Sie}, \citenamefont {Nyby},
  \citenamefont {Fan}, \citenamefont {Shen}, \citenamefont {Reid},
  \citenamefont {Hoffmann}, \citenamefont {Weathersby}, \citenamefont {Wen}
  \emph {et~al.}}]{luo2023ultrafast}%
  \BibitemOpen
  \bibfield  {author} {\bibinfo {author} {\bibfnamefont {D.}~\bibnamefont
  {Luo}}, \bibinfo {author} {\bibfnamefont {B.}~\bibnamefont {Zhang}}, \bibinfo
  {author} {\bibfnamefont {E.~J.}\ \bibnamefont {Sie}}, \bibinfo {author}
  {\bibfnamefont {C.~M.}\ \bibnamefont {Nyby}}, \bibinfo {author}
  {\bibfnamefont {Q.}~\bibnamefont {Fan}}, \bibinfo {author} {\bibfnamefont
  {X.}~\bibnamefont {Shen}}, \bibinfo {author} {\bibfnamefont {A.~H.}\
  \bibnamefont {Reid}}, \bibinfo {author} {\bibfnamefont {M.~C.}\ \bibnamefont
  {Hoffmann}}, \bibinfo {author} {\bibfnamefont {S.}~\bibnamefont
  {Weathersby}}, \bibinfo {author} {\bibfnamefont {J.}~\bibnamefont {Wen}},
  \emph {et~al.},\ }\bibfield  {title} {\bibinfo {title} {Ultrafast
  optomechanical strain in layered {GeS}},\ }\href
  {https://doi.org/https://doi.org/10.1021/acs.nanolett.2c05048} {\bibfield
  {journal} {\bibinfo  {journal} {Nano Lett.}\ }\textbf {\bibinfo {volume}
  {23}},\ \bibinfo {pages} {2287} (\bibinfo {year} {2023})}\BibitemShut
  {NoStop}%
\bibitem [{sup()}]{supmat}%
  \BibitemOpen
  \href@noop {} {}\bibinfo {note} {See Supplemental Material at
  \url{http://link.aps.org/ supplemental/xxxx/xxxx } for details on SnS growth,
  atomic force microscopy, polarized optical microscopy, second harmonic
  generation, pump-probe spectroscopy, static reflectance measurement, density
  functional photostriction calculations, thin-film interference modeling for
  reflectance and transient reflectivity, estimation of photo-induced
  temperature rise and thermal strain, estimation of strain corresponding to
  the extracted $\Delta n$, estimation of $\Delta n$ from DFT photostriction
  calculation, estimation of deformation potential, estimation of internal
  screening field in inverse piezoelectric effect, and supplementary Figs. S1 -
  S7, which includes Refs. \cite{zhang2025giant, kundys2010light,
  puri2024substrate, gonze2009abinit, perdew1996generalized,
  blochl1994projector, jollet2014generation, paillard2019photoinduced,
  verstraete2025abinit, kresse1993ab, kresse1996efficient,
  0polyanskiy2024refractiveindex, 0hecht2012optics, madelung2004semiconductors,
  ruello2015physical, osti_1197543, nye1985physical, fei2015giant,
  gu2002study}.}\BibitemShut {Stop}%
\bibitem [{\citenamefont {Wang}\ and\ \citenamefont
  {Qian}(2017{\natexlab{b}})}]{wang2017giant}%
  \BibitemOpen
  \bibfield  {author} {\bibinfo {author} {\bibfnamefont {H.}~\bibnamefont
  {Wang}}\ and\ \bibinfo {author} {\bibfnamefont {X.}~\bibnamefont {Qian}},\
  }\bibfield  {title} {\bibinfo {title} {Giant optical second harmonic
  generation in two-dimensional multiferroics},\ }\href
  {https://doi.org/10.1021/acs.nanolett.7b02268} {\bibfield  {journal}
  {\bibinfo  {journal} {Nano Lett.}\ }\textbf {\bibinfo {volume} {17}},\
  \bibinfo {pages} {5027} (\bibinfo {year} {2017}{\natexlab{b}})}\BibitemShut
  {NoStop}%
\bibitem [{\citenamefont {Yonemori}\ \emph {et~al.}(2021)\citenamefont
  {Yonemori}, \citenamefont {Dutta}, \citenamefont {Nagashio},\ and\
  \citenamefont {Wakabayashi}}]{yonemori2021thickness}%
  \BibitemOpen
  \bibfield  {author} {\bibinfo {author} {\bibfnamefont {I.}~\bibnamefont
  {Yonemori}}, \bibinfo {author} {\bibfnamefont {S.}~\bibnamefont {Dutta}},
  \bibinfo {author} {\bibfnamefont {K.}~\bibnamefont {Nagashio}},\ and\
  \bibinfo {author} {\bibfnamefont {K.}~\bibnamefont {Wakabayashi}},\
  }\bibfield  {title} {\bibinfo {title} {Thickness-dependent raman active modes
  of {SnS} thin films},\ }\href
  {https://doi.org/https://doi.org/10.1063/5.0062857} {\bibfield  {journal}
  {\bibinfo  {journal} {AIP Adv.}\ }\textbf {\bibinfo {volume} {11}},\ \bibinfo
  {pages} {095106} (\bibinfo {year} {2021})}\BibitemShut {NoStop}%
\bibitem [{\citenamefont {Shi}\ \emph {et~al.}(2016)\citenamefont {Shi},
  \citenamefont {Tang}, \citenamefont {Li}, \citenamefont {Liao}, \citenamefont
  {Wang}, \citenamefont {Ye},\ and\ \citenamefont {Xiong}}]{shi2016symmetry}%
  \BibitemOpen
  \bibfield  {author} {\bibinfo {author} {\bibfnamefont {P.-P.}\ \bibnamefont
  {Shi}}, \bibinfo {author} {\bibfnamefont {Y.-Y.}\ \bibnamefont {Tang}},
  \bibinfo {author} {\bibfnamefont {P.-F.}\ \bibnamefont {Li}}, \bibinfo
  {author} {\bibfnamefont {W.-Q.}\ \bibnamefont {Liao}}, \bibinfo {author}
  {\bibfnamefont {Z.-X.}\ \bibnamefont {Wang}}, \bibinfo {author}
  {\bibfnamefont {Q.}~\bibnamefont {Ye}},\ and\ \bibinfo {author}
  {\bibfnamefont {R.-G.}\ \bibnamefont {Xiong}},\ }\bibfield  {title} {\bibinfo
  {title} {Symmetry breaking in molecular ferroelectrics},\ }\href
  {https://doi.org/https://doi.org/10.1039/c5cs00308c} {\bibfield  {journal}
  {\bibinfo  {journal} {Chem. Soc. Rev.}\ }\textbf {\bibinfo {volume} {45}},\
  \bibinfo {pages} {3811} (\bibinfo {year} {2016})}\BibitemShut {NoStop}%
\bibitem [{\citenamefont {Moqbel}\ \emph {et~al.}(2024)\citenamefont {Moqbel},
  \citenamefont {Nanae}, \citenamefont {Kitamura}, \citenamefont {Lee},
  \citenamefont {Lan}, \citenamefont {Lee}, \citenamefont {Nagashio},\ and\
  \citenamefont {Lin}}]{moqbel2024giant}%
  \BibitemOpen
  \bibfield  {author} {\bibinfo {author} {\bibfnamefont {R.}~\bibnamefont
  {Moqbel}}, \bibinfo {author} {\bibfnamefont {R.}~\bibnamefont {Nanae}},
  \bibinfo {author} {\bibfnamefont {S.}~\bibnamefont {Kitamura}}, \bibinfo
  {author} {\bibfnamefont {M.-H.}\ \bibnamefont {Lee}}, \bibinfo {author}
  {\bibfnamefont {Y.-W.}\ \bibnamefont {Lan}}, \bibinfo {author} {\bibfnamefont
  {C.-C.}\ \bibnamefont {Lee}}, \bibinfo {author} {\bibfnamefont
  {K.}~\bibnamefont {Nagashio}},\ and\ \bibinfo {author} {\bibfnamefont
  {K.-H.}\ \bibnamefont {Lin}},\ }\bibfield  {title} {\bibinfo {title} {Giant
  second-order nonlinearity and anisotropy of large-sized few-layer {SnS} with
  ferroelectric stacking},\ }\href
  {https://doi.org/https://doi.org/10.1002/adom.202400355} {\bibfield
  {journal} {\bibinfo  {journal} {Adv. Opt. Mater.}\ }\textbf {\bibinfo
  {volume} {12}},\ \bibinfo {pages} {2400355} (\bibinfo {year}
  {2024})}\BibitemShut {NoStop}%
\bibitem [{\citenamefont {Sutter}\ \emph {et~al.}(2024)\citenamefont {Sutter},
  \citenamefont {Ghimire},\ and\ \citenamefont
  {Sutter}}]{sutter2024macroscopic}%
  \BibitemOpen
  \bibfield  {author} {\bibinfo {author} {\bibfnamefont {E.}~\bibnamefont
  {Sutter}}, \bibinfo {author} {\bibfnamefont {P.}~\bibnamefont {Ghimire}},\
  and\ \bibinfo {author} {\bibfnamefont {P.}~\bibnamefont {Sutter}},\
  }\bibfield  {title} {\bibinfo {title} {Macroscopic monochalcogenide van der
  waals ferroics: Growth, domain structures, and curie temperature},\ }\href
  {https://doi.org/https://doi.org/10.1021/jacs.4c11558} {\bibfield  {journal}
  {\bibinfo  {journal} {J. Am. Chem. Soc.}\ }\textbf {\bibinfo {volume}
  {146}},\ \bibinfo {pages} {31961} (\bibinfo {year} {2024})}\BibitemShut
  {NoStop}%
\bibitem [{\citenamefont {Mao}\ \emph {et~al.}(2023)\citenamefont {Mao},
  \citenamefont {Luo}, \citenamefont {Chiu}, \citenamefont {Shi}, \citenamefont
  {Ji}, \citenamefont {Pieshkov}, \citenamefont {Lin}, \citenamefont {Tang},
  \citenamefont {Akey}, \citenamefont {Gardener} \emph
  {et~al.}}]{mao2023giant}%
  \BibitemOpen
  \bibfield  {author} {\bibinfo {author} {\bibfnamefont {N.}~\bibnamefont
  {Mao}}, \bibinfo {author} {\bibfnamefont {Y.}~\bibnamefont {Luo}}, \bibinfo
  {author} {\bibfnamefont {M.-H.}\ \bibnamefont {Chiu}}, \bibinfo {author}
  {\bibfnamefont {C.}~\bibnamefont {Shi}}, \bibinfo {author} {\bibfnamefont
  {X.}~\bibnamefont {Ji}}, \bibinfo {author} {\bibfnamefont {T.~S.}\
  \bibnamefont {Pieshkov}}, \bibinfo {author} {\bibfnamefont {Y.}~\bibnamefont
  {Lin}}, \bibinfo {author} {\bibfnamefont {H.-L.}\ \bibnamefont {Tang}},
  \bibinfo {author} {\bibfnamefont {A.~J.}\ \bibnamefont {Akey}}, \bibinfo
  {author} {\bibfnamefont {J.~A.}\ \bibnamefont {Gardener}}, \emph {et~al.},\
  }\bibfield  {title} {\bibinfo {title} {Giant nonlinear optical response via
  coherent stacking of in-plane ferroelectric layers},\ }\href
  {https://doi.org/https://doi.org/10.1002/adma.202210894} {\bibfield
  {journal} {\bibinfo  {journal} {Adv. Mater.}\ }\textbf {\bibinfo {volume}
  {35}},\ \bibinfo {pages} {2210894} (\bibinfo {year} {2023})}\BibitemShut
  {NoStop}%
\bibitem [{\citenamefont {Sun}\ \emph {et~al.}(2022)\citenamefont {Sun},
  \citenamefont {Ma}, \citenamefont {Xia}, \citenamefont {Suo}, \citenamefont
  {Zhang}, \citenamefont {Zou}, \citenamefont {Lin}, \citenamefont {Zhang},
  \citenamefont {Guo},\ and\ \citenamefont {Ma}}]{sun2022dynamical}%
  \BibitemOpen
  \bibfield  {author} {\bibinfo {author} {\bibfnamefont {K.}~\bibnamefont
  {Sun}}, \bibinfo {author} {\bibfnamefont {H.}~\bibnamefont {Ma}}, \bibinfo
  {author} {\bibfnamefont {W.}~\bibnamefont {Xia}}, \bibinfo {author}
  {\bibfnamefont {P.}~\bibnamefont {Suo}}, \bibinfo {author} {\bibfnamefont
  {W.}~\bibnamefont {Zhang}}, \bibinfo {author} {\bibfnamefont
  {Y.}~\bibnamefont {Zou}}, \bibinfo {author} {\bibfnamefont {X.}~\bibnamefont
  {Lin}}, \bibinfo {author} {\bibfnamefont {S.}~\bibnamefont {Zhang}}, \bibinfo
  {author} {\bibfnamefont {Y.}~\bibnamefont {Guo}},\ and\ \bibinfo {author}
  {\bibfnamefont {G.}~\bibnamefont {Ma}},\ }\bibfield  {title} {\bibinfo
  {title} {Dynamical response of nonlinear optical anisotropy in a tin sulfide
  crystal under ultrafast photoexcitation},\ }\href
  {https://doi.org/https://doi.org/10.1021/acs.jpclett.2c02443} {\bibfield
  {journal} {\bibinfo  {journal} {J. Phys. Chem. Lett.}\ }\textbf {\bibinfo
  {volume} {13}},\ \bibinfo {pages} {9355} (\bibinfo {year}
  {2022})}\BibitemShut {NoStop}%
\bibitem [{\citenamefont {Hoang}\ \emph {et~al.}(2025)\citenamefont {Hoang},
  \citenamefont {Pesquera}, \citenamefont {Hinsley}, \citenamefont {Carley},
  \citenamefont {Mercadier}, \citenamefont {Teichmann}, \citenamefont
  {Unterleutner}, \citenamefont {Knez}, \citenamefont {Dienstleder},
  \citenamefont {Ganguly} \emph {et~al.}}]{hoang2025ultrafast}%
  \BibitemOpen
  \bibfield  {author} {\bibinfo {author} {\bibfnamefont {L.~P.}\ \bibnamefont
  {Hoang}}, \bibinfo {author} {\bibfnamefont {D.}~\bibnamefont {Pesquera}},
  \bibinfo {author} {\bibfnamefont {G.~N.}\ \bibnamefont {Hinsley}}, \bibinfo
  {author} {\bibfnamefont {R.}~\bibnamefont {Carley}}, \bibinfo {author}
  {\bibfnamefont {L.}~\bibnamefont {Mercadier}}, \bibinfo {author}
  {\bibfnamefont {M.}~\bibnamefont {Teichmann}}, \bibinfo {author}
  {\bibfnamefont {E.~M.}\ \bibnamefont {Unterleutner}}, \bibinfo {author}
  {\bibfnamefont {D.}~\bibnamefont {Knez}}, \bibinfo {author} {\bibfnamefont
  {M.}~\bibnamefont {Dienstleder}}, \bibinfo {author} {\bibfnamefont
  {S.}~\bibnamefont {Ganguly}}, \emph {et~al.},\ }\bibfield  {title} {\bibinfo
  {title} {Ultrafast decoupling of polarization and strain in ferroelectric
  {BaTiO$_3$}},\ }\href
  {https://doi.org/https://doi.org/10.1038/s41467-025-63045-6} {\bibfield
  {journal} {\bibinfo  {journal} {Nat. Commun.}\ }\textbf {\bibinfo {volume}
  {16}},\ \bibinfo {pages} {7966} (\bibinfo {year} {2025})}\BibitemShut
  {NoStop}%
\bibitem [{\citenamefont {Wen}\ \emph {et~al.}(2013)\citenamefont {Wen},
  \citenamefont {Chen}, \citenamefont {Cosgriff}, \citenamefont {Walko},
  \citenamefont {Lee}, \citenamefont {Adamo}, \citenamefont {Schaller},
  \citenamefont {Ihlefeld}, \citenamefont {Dufresne}, \citenamefont {Schlom}
  \emph {et~al.}}]{wen2013electronic}%
  \BibitemOpen
  \bibfield  {author} {\bibinfo {author} {\bibfnamefont {H.}~\bibnamefont
  {Wen}}, \bibinfo {author} {\bibfnamefont {P.}~\bibnamefont {Chen}}, \bibinfo
  {author} {\bibfnamefont {M.~P.}\ \bibnamefont {Cosgriff}}, \bibinfo {author}
  {\bibfnamefont {D.~A.}\ \bibnamefont {Walko}}, \bibinfo {author}
  {\bibfnamefont {J.~H.}\ \bibnamefont {Lee}}, \bibinfo {author} {\bibfnamefont
  {C.}~\bibnamefont {Adamo}}, \bibinfo {author} {\bibfnamefont {R.~D.}\
  \bibnamefont {Schaller}}, \bibinfo {author} {\bibfnamefont {J.~F.}\
  \bibnamefont {Ihlefeld}}, \bibinfo {author} {\bibfnamefont {E.~M.}\
  \bibnamefont {Dufresne}}, \bibinfo {author} {\bibfnamefont {D.~G.}\
  \bibnamefont {Schlom}}, \emph {et~al.},\ }\bibfield  {title} {\bibinfo
  {title} {Electronic origin of ultrafast photoinduced strain in {BiFeO$_3$}},\
  }\href {https://doi.org/https://doi.org/10.1103/PhysRevLett.110.037601}
  {\bibfield  {journal} {\bibinfo  {journal} {Phys. Rev. Lett.}\ }\textbf
  {\bibinfo {volume} {110}},\ \bibinfo {pages} {037601} (\bibinfo {year}
  {2013})}\BibitemShut {NoStop}%
\bibitem [{\citenamefont {Paillard}\ \emph {et~al.}(2016)\citenamefont
  {Paillard}, \citenamefont {Xu}, \citenamefont {Dkhil}, \citenamefont
  {Geneste},\ and\ \citenamefont {Bellaiche}}]{paillard2016photostriction}%
  \BibitemOpen
  \bibfield  {author} {\bibinfo {author} {\bibfnamefont {C.}~\bibnamefont
  {Paillard}}, \bibinfo {author} {\bibfnamefont {B.}~\bibnamefont {Xu}},
  \bibinfo {author} {\bibfnamefont {B.}~\bibnamefont {Dkhil}}, \bibinfo
  {author} {\bibfnamefont {G.}~\bibnamefont {Geneste}},\ and\ \bibinfo {author}
  {\bibfnamefont {L.}~\bibnamefont {Bellaiche}},\ }\bibfield  {title} {\bibinfo
  {title} {Photostriction in ferroelectrics from density functional theory},\
  }\href {https://doi.org/https://doi.org/10.1103/PhysRevLett.116.247401}
  {\bibfield  {journal} {\bibinfo  {journal} {Phys. Rev. Lett.}\ }\textbf
  {\bibinfo {volume} {116}},\ \bibinfo {pages} {247401} (\bibinfo {year}
  {2016})}\BibitemShut {NoStop}%
\bibitem [{\citenamefont {Paillard}\ \emph {et~al.}(2017)\citenamefont
  {Paillard}, \citenamefont {Prosandeev},\ and\ \citenamefont
  {Bellaiche}}]{paillard2017ab}%
  \BibitemOpen
  \bibfield  {author} {\bibinfo {author} {\bibfnamefont {C.}~\bibnamefont
  {Paillard}}, \bibinfo {author} {\bibfnamefont {S.}~\bibnamefont
  {Prosandeev}},\ and\ \bibinfo {author} {\bibfnamefont {L.}~\bibnamefont
  {Bellaiche}},\ }\bibfield  {title} {\bibinfo {title} {Ab initio approach to
  photostriction in classical ferroelectric materials},\ }\href
  {https://doi.org/https://doi.org/10.1103/PhysRevB.96.045205} {\bibfield
  {journal} {\bibinfo  {journal} {Phys. Rev. B}\ }\textbf {\bibinfo {volume}
  {96}},\ \bibinfo {pages} {045205} (\bibinfo {year} {2017})}\BibitemShut
  {NoStop}%
\bibitem [{\citenamefont {Dansou}\ \emph {et~al.}(2025)\citenamefont {Dansou},
  \citenamefont {Paillard},\ and\ \citenamefont
  {Bellaiche}}]{dansou2025photoinduced}%
  \BibitemOpen
  \bibfield  {author} {\bibinfo {author} {\bibfnamefont {C.}~\bibnamefont
  {Dansou}}, \bibinfo {author} {\bibfnamefont {C.}~\bibnamefont {Paillard}},\
  and\ \bibinfo {author} {\bibfnamefont {L.}~\bibnamefont {Bellaiche}},\
  }\bibfield  {title} {\bibinfo {title} {Photoinduced phase transitions and
  lattice deformation in two-dimensional {NbOX$_2$ (X= Cl, Br, I)}},\ }\href
  {https://doi.org/https://doi.org/10.1103/sbcn-pkv4} {\bibfield  {journal}
  {\bibinfo  {journal} {Phys. Rev. Mater.}\ }\textbf {\bibinfo {volume} {9}},\
  \bibinfo {pages} {114001} (\bibinfo {year} {2025})}\BibitemShut {NoStop}%
\bibitem [{\citenamefont {Zhang}\ \emph {et~al.}(2025)\citenamefont {Zhang},
  \citenamefont {Nagashree}, \citenamefont {Webster}, \citenamefont {Edwards},
  \citenamefont {Rajendra}, \citenamefont {Kulkarni}, \citenamefont {Yousaf},
  \citenamefont {Zhang}, \citenamefont {Gruverman}, \citenamefont {Seidel}
  \emph {et~al.}}]{zhang2025giant}%
  \BibitemOpen
  \bibfield  {author} {\bibinfo {author} {\bibfnamefont {H.}~\bibnamefont
  {Zhang}}, \bibinfo {author} {\bibfnamefont {M.}~\bibnamefont {Nagashree}},
  \bibinfo {author} {\bibfnamefont {R.}~\bibnamefont {Webster}}, \bibinfo
  {author} {\bibfnamefont {J.}~\bibnamefont {Edwards}}, \bibinfo {author}
  {\bibfnamefont {B.}~\bibnamefont {Rajendra}}, \bibinfo {author}
  {\bibfnamefont {S.}~\bibnamefont {Kulkarni}}, \bibinfo {author}
  {\bibfnamefont {T.}~\bibnamefont {Yousaf}}, \bibinfo {author} {\bibfnamefont
  {D.}~\bibnamefont {Zhang}}, \bibinfo {author} {\bibfnamefont
  {A.}~\bibnamefont {Gruverman}}, \bibinfo {author} {\bibfnamefont
  {J.}~\bibnamefont {Seidel}}, \emph {et~al.},\ }\bibfield  {title} {\bibinfo
  {title} {Giant photostriction and optically modulated ferroelectricity in
  {BiFeO$_3$}},\ }\href
  {https://doi.org/https://doi.org/10.1021/acsnano.5c05203} {\bibfield
  {journal} {\bibinfo  {journal} {ACS Nano}\ }\textbf {\bibinfo {volume}
  {19}},\ \bibinfo {pages} {33780} (\bibinfo {year} {2025})}\BibitemShut
  {NoStop}%
\bibitem [{\citenamefont {Kundys}\ \emph {et~al.}(2010)\citenamefont {Kundys},
  \citenamefont {Viret}, \citenamefont {Colson},\ and\ \citenamefont
  {Kundys}}]{kundys2010light}%
  \BibitemOpen
  \bibfield  {author} {\bibinfo {author} {\bibfnamefont {B.}~\bibnamefont
  {Kundys}}, \bibinfo {author} {\bibfnamefont {M.}~\bibnamefont {Viret}},
  \bibinfo {author} {\bibfnamefont {D.}~\bibnamefont {Colson}},\ and\ \bibinfo
  {author} {\bibfnamefont {D.~O.}\ \bibnamefont {Kundys}},\ }\bibfield  {title}
  {\bibinfo {title} {Light-induced size changes in {BiFeO$_3$} crystals},\
  }\href {https://doi.org/https://doi.org/10.1038/nmat2807} {\bibfield
  {journal} {\bibinfo  {journal} {Nat. Mater.}\ }\textbf {\bibinfo {volume}
  {9}},\ \bibinfo {pages} {803} (\bibinfo {year} {2010})}\BibitemShut {NoStop}%
\bibitem [{\citenamefont {Puri}\ \emph {et~al.}(2024)\citenamefont {Puri},
  \citenamefont {Patel}, \citenamefont {Cabellos}, \citenamefont
  {Rosas-Hernandez}, \citenamefont {Reynolds}, \citenamefont {Churchill},
  \citenamefont {Barraza-Lopez}, \citenamefont {Mendoza},\ and\ \citenamefont
  {Nakamura}}]{puri2024substrate}%
  \BibitemOpen
  \bibfield  {author} {\bibinfo {author} {\bibfnamefont {S.}~\bibnamefont
  {Puri}}, \bibinfo {author} {\bibfnamefont {S.}~\bibnamefont {Patel}},
  \bibinfo {author} {\bibfnamefont {J.~L.}\ \bibnamefont {Cabellos}}, \bibinfo
  {author} {\bibfnamefont {L.~E.}\ \bibnamefont {Rosas-Hernandez}}, \bibinfo
  {author} {\bibfnamefont {K.}~\bibnamefont {Reynolds}}, \bibinfo {author}
  {\bibfnamefont {H.~O.}\ \bibnamefont {Churchill}}, \bibinfo {author}
  {\bibfnamefont {S.}~\bibnamefont {Barraza-Lopez}}, \bibinfo {author}
  {\bibfnamefont {B.~S.}\ \bibnamefont {Mendoza}},\ and\ \bibinfo {author}
  {\bibfnamefont {H.}~\bibnamefont {Nakamura}},\ }\bibfield  {title} {\bibinfo
  {title} {Substrate interference and strain in the second-harmonic generation
  from {MoSe}$_2$ monolayers},\ }\href
  {https://doi.org/https://doi.org/10.1021/acs.nanolett.4c03880} {\bibfield
  {journal} {\bibinfo  {journal} {Nano Lett.}\ }\textbf {\bibinfo {volume}
  {24}},\ \bibinfo {pages} {13061} (\bibinfo {year} {2024})}\BibitemShut
  {NoStop}%
\bibitem [{\citenamefont {Gonze}\ \emph {et~al.}(2009)\citenamefont {Gonze},
  \citenamefont {Amadon}, \citenamefont {Anglade}, \citenamefont {Beuken},
  \citenamefont {Bottin}, \citenamefont {Boulanger}, \citenamefont {Bruneval},
  \citenamefont {Caliste}, \citenamefont {Caracas}, \citenamefont {Cote} \emph
  {et~al.}}]{gonze2009abinit}%
  \BibitemOpen
  \bibfield  {author} {\bibinfo {author} {\bibfnamefont {X.}~\bibnamefont
  {Gonze}}, \bibinfo {author} {\bibfnamefont {B.}~\bibnamefont {Amadon}},
  \bibinfo {author} {\bibfnamefont {P.}~\bibnamefont {Anglade}}, \bibinfo
  {author} {\bibfnamefont {J.-M.}\ \bibnamefont {Beuken}}, \bibinfo {author}
  {\bibfnamefont {F.}~\bibnamefont {Bottin}}, \bibinfo {author} {\bibfnamefont
  {P.}~\bibnamefont {Boulanger}}, \bibinfo {author} {\bibfnamefont
  {F.}~\bibnamefont {Bruneval}}, \bibinfo {author} {\bibfnamefont
  {D.}~\bibnamefont {Caliste}}, \bibinfo {author} {\bibfnamefont
  {R.}~\bibnamefont {Caracas}}, \bibinfo {author} {\bibfnamefont
  {M.}~\bibnamefont {Cote}}, \emph {et~al.},\ }\bibfield  {title} {\bibinfo
  {title} {Abinit: First-principles approach to material and nanosystem
  properties},\ }\href
  {https://doi.org/https://doi.org/10.1016/j.cpc.2009.07.007} {\bibfield
  {journal} {\bibinfo  {journal} {Comput. Phys. Commun.}\ }\textbf {\bibinfo
  {volume} {180}},\ \bibinfo {pages} {2582} (\bibinfo {year}
  {2009})}\BibitemShut {NoStop}%
\bibitem [{\citenamefont {Perdew}\ \emph {et~al.}(1996)\citenamefont {Perdew},
  \citenamefont {Burke},\ and\ \citenamefont
  {Ernzerhof}}]{perdew1996generalized}%
  \BibitemOpen
  \bibfield  {author} {\bibinfo {author} {\bibfnamefont {J.~P.}\ \bibnamefont
  {Perdew}}, \bibinfo {author} {\bibfnamefont {K.}~\bibnamefont {Burke}},\ and\
  \bibinfo {author} {\bibfnamefont {M.}~\bibnamefont {Ernzerhof}},\ }\bibfield
  {title} {\bibinfo {title} {Generalized gradient approximation made simple},\
  }\href {https://doi.org/10.1103/PhysRevLett.77.3865} {\bibfield  {journal}
  {\bibinfo  {journal} {Phys. Rev. Lett.}\ }\textbf {\bibinfo {volume} {77}},\
  \bibinfo {pages} {3865} (\bibinfo {year} {1996})}\BibitemShut {NoStop}%
\bibitem [{\citenamefont {Bl{\"o}chl}(1994)}]{blochl1994projector}%
  \BibitemOpen
  \bibfield  {author} {\bibinfo {author} {\bibfnamefont {P.~E.}\ \bibnamefont
  {Bl{\"o}chl}},\ }\bibfield  {title} {\bibinfo {title} {Projector
  augmented-wave method},\ }\href {https://doi.org/10.1103/PhysRevB.50.17953}
  {\bibfield  {journal} {\bibinfo  {journal} {Phys. Rev. B}\ }\textbf {\bibinfo
  {volume} {50}},\ \bibinfo {pages} {17953} (\bibinfo {year}
  {1994})}\BibitemShut {NoStop}%
\bibitem [{\citenamefont {Jollet}\ \emph {et~al.}(2014)\citenamefont {Jollet},
  \citenamefont {Torrent},\ and\ \citenamefont
  {Holzwarth}}]{jollet2014generation}%
  \BibitemOpen
  \bibfield  {author} {\bibinfo {author} {\bibfnamefont {F.}~\bibnamefont
  {Jollet}}, \bibinfo {author} {\bibfnamefont {M.}~\bibnamefont {Torrent}},\
  and\ \bibinfo {author} {\bibfnamefont {N.}~\bibnamefont {Holzwarth}},\
  }\bibfield  {title} {\bibinfo {title} {Generation of projector augmented-wave
  atomic data: A 71 element validated table in the {XML} format},\ }\href
  {https://doi.org/https://doi.org/10.1016/j.cpc.2013.12.023} {\bibfield
  {journal} {\bibinfo  {journal} {Comput. Phys. Commun.}\ }\textbf {\bibinfo
  {volume} {185}},\ \bibinfo {pages} {1246} (\bibinfo {year}
  {2014})}\BibitemShut {NoStop}%
\bibitem [{\citenamefont {Paillard}\ \emph {et~al.}(2019)\citenamefont
  {Paillard}, \citenamefont {Torun}, \citenamefont {Wirtz}, \citenamefont
  {{\'I}{\~n}iguez},\ and\ \citenamefont
  {Bellaiche}}]{paillard2019photoinduced}%
  \BibitemOpen
  \bibfield  {author} {\bibinfo {author} {\bibfnamefont {C.}~\bibnamefont
  {Paillard}}, \bibinfo {author} {\bibfnamefont {E.}~\bibnamefont {Torun}},
  \bibinfo {author} {\bibfnamefont {L.}~\bibnamefont {Wirtz}}, \bibinfo
  {author} {\bibfnamefont {J.}~\bibnamefont {{\'I}{\~n}iguez}},\ and\ \bibinfo
  {author} {\bibfnamefont {L.}~\bibnamefont {Bellaiche}},\ }\bibfield  {title}
  {\bibinfo {title} {Photoinduced phase transitions in ferroelectrics},\ }\href
  {https://doi.org/https://doi.org/10.1103/PhysRevLett.123.087601} {\bibfield
  {journal} {\bibinfo  {journal} {Phys. Rev. Lett.}\ }\textbf {\bibinfo
  {volume} {123}},\ \bibinfo {pages} {087601} (\bibinfo {year}
  {2019})}\BibitemShut {NoStop}%
\bibitem [{\citenamefont {Verstraete}\ \emph {et~al.}(2025)\citenamefont
  {Verstraete}, \citenamefont {Abreu}, \citenamefont {Allemand}, \citenamefont
  {Amadon}, \citenamefont {Antonius}, \citenamefont {Azizi}, \citenamefont
  {Baguet} \emph {et~al.}}]{verstraete2025abinit}%
  \BibitemOpen
  \bibfield  {author} {\bibinfo {author} {\bibfnamefont {M.~J.}\ \bibnamefont
  {Verstraete}}, \bibinfo {author} {\bibfnamefont {J.}~\bibnamefont {Abreu}},
  \bibinfo {author} {\bibfnamefont {G.~E.}\ \bibnamefont {Allemand}}, \bibinfo
  {author} {\bibfnamefont {B.}~\bibnamefont {Amadon}}, \bibinfo {author}
  {\bibfnamefont {G.}~\bibnamefont {Antonius}}, \bibinfo {author}
  {\bibfnamefont {M.}~\bibnamefont {Azizi}}, \bibinfo {author} {\bibfnamefont
  {L.}~\bibnamefont {Baguet}}, \emph {et~al.},\ }\bibfield  {title} {\bibinfo
  {title} {Abinit 2025: New capabilities for the predictive modeling of solids
  and nanomaterials},\ }\href
  {https://doi.org/https://doi.org/10.1063/5.0288278} {\bibfield  {journal}
  {\bibinfo  {journal} {J. Chem. Phys.}\ }\textbf {\bibinfo {volume} {163}},\
  \bibinfo {pages} {164126} (\bibinfo {year} {2025})}\BibitemShut {NoStop}%
\bibitem [{\citenamefont {Kresse}\ and\ \citenamefont
  {Hafner}(1993)}]{kresse1993ab}%
  \BibitemOpen
  \bibfield  {author} {\bibinfo {author} {\bibfnamefont {G.}~\bibnamefont
  {Kresse}}\ and\ \bibinfo {author} {\bibfnamefont {J.}~\bibnamefont
  {Hafner}},\ }\bibfield  {title} {\bibinfo {title} {Ab initio molecular
  dynamics for liquid metals},\ }\href
  {https://doi.org/https://doi.org/10.1103/PhysRevB.47.558} {\bibfield
  {journal} {\bibinfo  {journal} {Phys. Rev. B}\ }\textbf {\bibinfo {volume}
  {47}},\ \bibinfo {pages} {558} (\bibinfo {year} {1993})}\BibitemShut
  {NoStop}%
\bibitem [{\citenamefont {Kresse}\ and\ \citenamefont
  {Furthm{\"u}ller}(1996)}]{kresse1996efficient}%
  \BibitemOpen
  \bibfield  {author} {\bibinfo {author} {\bibfnamefont {G.}~\bibnamefont
  {Kresse}}\ and\ \bibinfo {author} {\bibfnamefont {J.}~\bibnamefont
  {Furthm{\"u}ller}},\ }\bibfield  {title} {\bibinfo {title} {Efficient
  iterative schemes for ab initio total-energy calculations using a plane-wave
  basis set},\ }\href
  {https://doi.org/https://doi.org/10.1103/PhysRevB.54.11169} {\bibfield
  {journal} {\bibinfo  {journal} {Phys. Rev. B}\ }\textbf {\bibinfo {volume}
  {54}},\ \bibinfo {pages} {11169} (\bibinfo {year} {1996})}\BibitemShut
  {NoStop}%
\bibitem [{\citenamefont {Polyanskiy}(2024)}]{0polyanskiy2024refractiveindex}%
  \BibitemOpen
  \bibfield  {author} {\bibinfo {author} {\bibfnamefont {M.~N.}\ \bibnamefont
  {Polyanskiy}},\ }\bibfield  {title} {\bibinfo {title} {Refractiveindex. info
  database of optical constants},\ }\href
  {https://doi.org/https://doi.org/10.1038/s41597-023-02898-2} {\bibfield
  {journal} {\bibinfo  {journal} {Sci. Data}\ }\textbf {\bibinfo {volume}
  {11}},\ \bibinfo {pages} {94} (\bibinfo {year} {2024})}\BibitemShut {NoStop}%
\bibitem [{\citenamefont {Hecht}(2012)}]{0hecht2012optics}%
  \BibitemOpen
  \bibfield  {author} {\bibinfo {author} {\bibfnamefont {E.}~\bibnamefont
  {Hecht}},\ }\href@noop {} {\emph {\bibinfo {title} {Optics}}}\ (\bibinfo
  {publisher} {Pearson Education India},\ \bibinfo {year} {2012})\BibitemShut
  {NoStop}%
\bibitem [{\citenamefont {Madelung}(2004)}]{madelung2004semiconductors}%
  \BibitemOpen
  \bibfield  {author} {\bibinfo {author} {\bibfnamefont {O.}~\bibnamefont
  {Madelung}},\ }\href@noop {} {\emph {\bibinfo {title} {Semiconductors: data
  handbook}}}\ (\bibinfo  {publisher} {Springer Science \& Business Media},\
  \bibinfo {year} {2004})\BibitemShut {NoStop}%
\bibitem [{\citenamefont {Ruello}\ and\ \citenamefont
  {Gusev}(2015)}]{ruello2015physical}%
  \BibitemOpen
  \bibfield  {author} {\bibinfo {author} {\bibfnamefont {P.}~\bibnamefont
  {Ruello}}\ and\ \bibinfo {author} {\bibfnamefont {V.~E.}\ \bibnamefont
  {Gusev}},\ }\bibfield  {title} {\bibinfo {title} {Physical mechanisms of
  coherent acoustic phonons generation by ultrafast laser action},\ }\href
  {https://doi.org/https://doi.org/10.1016/j.ultras.2014.06.004} {\bibfield
  {journal} {\bibinfo  {journal} {Ultrasonics}\ }\textbf {\bibinfo {volume}
  {56}},\ \bibinfo {pages} {21} (\bibinfo {year} {2015})}\BibitemShut {NoStop}%
\bibitem [{ost(2020)}]{osti_1197543}%
  \BibitemOpen
  \bibfield  {title} {\bibinfo {title} {Materials data on {S}n{S} by
  {M}aterials {P}roject},\ }\href@noop {} {\  (\bibinfo {year}
  {2020})}\BibitemShut {NoStop}%
\bibitem [{\citenamefont {Nye}(1985)}]{nye1985physical}%
  \BibitemOpen
  \bibfield  {author} {\bibinfo {author} {\bibfnamefont {J.~F.}\ \bibnamefont
  {Nye}},\ }\href@noop {} {\emph {\bibinfo {title} {Physical properties of
  crystals: their representation by tensors and matrices}}}\ (\bibinfo
  {publisher} {Oxford university press},\ \bibinfo {year} {1985})\BibitemShut
  {NoStop}%
\bibitem [{\citenamefont {Fei}\ \emph {et~al.}(2015)\citenamefont {Fei},
  \citenamefont {Li}, \citenamefont {Li},\ and\ \citenamefont
  {Yang}}]{fei2015giant}%
  \BibitemOpen
  \bibfield  {author} {\bibinfo {author} {\bibfnamefont {R.}~\bibnamefont
  {Fei}}, \bibinfo {author} {\bibfnamefont {W.}~\bibnamefont {Li}}, \bibinfo
  {author} {\bibfnamefont {J.}~\bibnamefont {Li}},\ and\ \bibinfo {author}
  {\bibfnamefont {L.}~\bibnamefont {Yang}},\ }\bibfield  {title} {\bibinfo
  {title} {Giant piezoelectricity of monolayer group {IV} monochalcogenides:
  {SnSe}, {SnS}, {GeSe}, and {GeS}},\ }\href
  {https://doi.org/https://doi.org/10.1063/1.4934750} {\bibfield  {journal}
  {\bibinfo  {journal} {Appl. Phys. Lett.}\ }\textbf {\bibinfo {volume}
  {107}},\ \bibinfo {pages} {173104} (\bibinfo {year} {2015})}\BibitemShut
  {NoStop}%
\bibitem [{\citenamefont {Gu}\ \emph {et~al.}(2002)\citenamefont {Gu},
  \citenamefont {Tani}, \citenamefont {Kono}, \citenamefont {Sakai},\ and\
  \citenamefont {Zhang}}]{gu2002study}%
  \BibitemOpen
  \bibfield  {author} {\bibinfo {author} {\bibfnamefont {P.}~\bibnamefont
  {Gu}}, \bibinfo {author} {\bibfnamefont {M.}~\bibnamefont {Tani}}, \bibinfo
  {author} {\bibfnamefont {S.}~\bibnamefont {Kono}}, \bibinfo {author}
  {\bibfnamefont {K.}~\bibnamefont {Sakai}},\ and\ \bibinfo {author}
  {\bibfnamefont {X.-C.}\ \bibnamefont {Zhang}},\ }\bibfield  {title} {\bibinfo
  {title} {Study of terahertz radiation from inas and insb},\ }\href
  {https://doi.org/10.1063/1.1465507} {\bibfield  {journal} {\bibinfo
  {journal} {J. Appl. Phys.}\ }\textbf {\bibinfo {volume} {91}},\ \bibinfo
  {pages} {5533} (\bibinfo {year} {2002})}\BibitemShut {NoStop}%
\end{thebibliography}

\begin{thebibliography}{24}%
\makeatletter
\providecommand \@ifxundefined [1]{%
 \@ifx{#1\undefined}
}%
\providecommand \@ifnum [1]{%
 \ifnum #1\expandafter \@firstoftwo
 \else \expandafter \@secondoftwo
 \fi
}%
\providecommand \@ifx [1]{%
 \ifx #1\expandafter \@firstoftwo
 \else \expandafter \@secondoftwo
 \fi
}%
\providecommand \natexlab [1]{#1}%
\providecommand \enquote  [1]{``#1''}%
\providecommand \bibnamefont  [1]{#1}%
\providecommand \bibfnamefont [1]{#1}%
\providecommand \citenamefont [1]{#1}%
\providecommand \href@noop [0]{\@secondoftwo}%
\providecommand \href [0]{\begingroup \@sanitize@url \@href}%
\providecommand \@href[1]{\@@startlink{#1}\@@href}%
\providecommand \@@href[1]{\endgroup#1\@@endlink}%
\providecommand \@sanitize@url [0]{\catcode `\\12\catcode `\$12\catcode
  `\&12\catcode `\#12\catcode `\^12\catcode `\_12\catcode `\%12\relax}%
\providecommand \@@startlink[1]{}%
\providecommand \@@endlink[0]{}%
\providecommand \url  [0]{\begingroup\@sanitize@url \@url }%
\providecommand \@url [1]{\endgroup\@href {#1}{\urlprefix }}%
\providecommand \urlprefix  [0]{URL }%
\providecommand \Eprint [0]{\href }%
\providecommand \doibase [0]{https://doi.org/}%
\providecommand \selectlanguage [0]{\@gobble}%
\providecommand \bibinfo  [0]{\@secondoftwo}%
\providecommand \bibfield  [0]{\@secondoftwo}%
\providecommand \translation [1]{[#1]}%
\providecommand \BibitemOpen [0]{}%
\providecommand \bibitemStop [0]{}%
\providecommand \bibitemNoStop [0]{.\EOS\space}%
\providecommand \EOS [0]{\spacefactor3000\relax}%
\providecommand \BibitemShut  [1]{\csname bibitem#1\endcsname}%
\let\auto@bib@innerbib\@empty
%</preamble>
\bibitem [{\citenamefont {Haleoot}\ \emph {et~al.}(2017)\citenamefont
  {Haleoot}, \citenamefont {Paillard}, \citenamefont {Kaloni}, \citenamefont
  {Mehboudi}, \citenamefont {Xu}, \citenamefont {Bellaiche},\ and\
  \citenamefont {Barraza-Lopez}}]{SM_haleoot2017photostrictive}%
  \BibitemOpen
  \bibfield  {author} {\bibinfo {author} {\bibfnamefont {R.}~\bibnamefont
  {Haleoot}}, \bibinfo {author} {\bibfnamefont {C.}~\bibnamefont {Paillard}},
  \bibinfo {author} {\bibfnamefont {T.~P.}\ \bibnamefont {Kaloni}}, \bibinfo
  {author} {\bibfnamefont {M.}~\bibnamefont {Mehboudi}}, \bibinfo {author}
  {\bibfnamefont {B.}~\bibnamefont {Xu}}, \bibinfo {author} {\bibfnamefont
  {L.}~\bibnamefont {Bellaiche}},\ and\ \bibinfo {author} {\bibfnamefont
  {S.}~\bibnamefont {Barraza-Lopez}},\ }\bibfield  {title} {\bibinfo {title}
  {Photostrictive two-dimensional materials in the monochalcogenide family},\
  }\href {https://doi.org/https://doi.org/10.1103/PhysRevLett.118.227401}
  {\bibfield  {journal} {\bibinfo  {journal} {Phys. Rev. Lett.}\ }\textbf
  {\bibinfo {volume} {118}},\ \bibinfo {pages} {227401} (\bibinfo {year}
  {2017})}\BibitemShut {NoStop}%
\bibitem [{\citenamefont {Luo}\ \emph {et~al.}(2023)\citenamefont {Luo},
  \citenamefont {Zhang}, \citenamefont {Sie}, \citenamefont {Nyby},
  \citenamefont {Fan}, \citenamefont {Shen}, \citenamefont {Reid},
  \citenamefont {Hoffmann}, \citenamefont {Weathersby}, \citenamefont {Wen}
  \emph {et~al.}}]{SM_luo2023ultrafast}%
  \BibitemOpen
  \bibfield  {author} {\bibinfo {author} {\bibfnamefont {D.}~\bibnamefont
  {Luo}}, \bibinfo {author} {\bibfnamefont {B.}~\bibnamefont {Zhang}}, \bibinfo
  {author} {\bibfnamefont {E.~J.}\ \bibnamefont {Sie}}, \bibinfo {author}
  {\bibfnamefont {C.~M.}\ \bibnamefont {Nyby}}, \bibinfo {author}
  {\bibfnamefont {Q.}~\bibnamefont {Fan}}, \bibinfo {author} {\bibfnamefont
  {X.}~\bibnamefont {Shen}}, \bibinfo {author} {\bibfnamefont {A.~H.}\
  \bibnamefont {Reid}}, \bibinfo {author} {\bibfnamefont {M.~C.}\ \bibnamefont
  {Hoffmann}}, \bibinfo {author} {\bibfnamefont {S.}~\bibnamefont
  {Weathersby}}, \bibinfo {author} {\bibfnamefont {J.}~\bibnamefont {Wen}},
  \emph {et~al.},\ }\bibfield  {title} {\bibinfo {title} {Ultrafast
  optomechanical strain in layered {GeS}},\ }\href
  {https://doi.org/https://doi.org/10.1021/acs.nanolett.2c05048} {\bibfield
  {journal} {\bibinfo  {journal} {Nano Lett.}\ }\textbf {\bibinfo {volume}
  {23}},\ \bibinfo {pages} {2287} (\bibinfo {year} {2023})}\BibitemShut
  {NoStop}%
\bibitem [{\citenamefont {Zhang}\ \emph {et~al.}(2025)\citenamefont {Zhang},
  \citenamefont {Nagashree}, \citenamefont {Webster}, \citenamefont {Edwards},
  \citenamefont {Rajendra}, \citenamefont {Kulkarni}, \citenamefont {Yousaf},
  \citenamefont {Zhang}, \citenamefont {Gruverman}, \citenamefont {Seidel}
  \emph {et~al.}}]{SM_zhang2025giant}%
  \BibitemOpen
  \bibfield  {author} {\bibinfo {author} {\bibfnamefont {H.}~\bibnamefont
  {Zhang}}, \bibinfo {author} {\bibfnamefont {M.}~\bibnamefont {Nagashree}},
  \bibinfo {author} {\bibfnamefont {R.}~\bibnamefont {Webster}}, \bibinfo
  {author} {\bibfnamefont {J.}~\bibnamefont {Edwards}}, \bibinfo {author}
  {\bibfnamefont {B.}~\bibnamefont {Rajendra}}, \bibinfo {author}
  {\bibfnamefont {S.}~\bibnamefont {Kulkarni}}, \bibinfo {author}
  {\bibfnamefont {T.}~\bibnamefont {Yousaf}}, \bibinfo {author} {\bibfnamefont
  {D.}~\bibnamefont {Zhang}}, \bibinfo {author} {\bibfnamefont
  {A.}~\bibnamefont {Gruverman}}, \bibinfo {author} {\bibfnamefont
  {J.}~\bibnamefont {Seidel}}, \emph {et~al.},\ }\bibfield  {title} {\bibinfo
  {title} {Giant photostriction and optically modulated ferroelectricity in
  {BiFeO$_3$}},\ }\href
  {https://doi.org/https://doi.org/10.1021/acsnano.5c05203} {\bibfield
  {journal} {\bibinfo  {journal} {ACS Nano}\ }\textbf {\bibinfo {volume}
  {19}},\ \bibinfo {pages} {33780} (\bibinfo {year} {2025})}\BibitemShut
  {NoStop}%
\bibitem [{\citenamefont {Wen}\ \emph {et~al.}(2013)\citenamefont {Wen},
  \citenamefont {Chen}, \citenamefont {Cosgriff}, \citenamefont {Walko},
  \citenamefont {Lee}, \citenamefont {Adamo}, \citenamefont {Schaller},
  \citenamefont {Ihlefeld}, \citenamefont {Dufresne}, \citenamefont {Schlom}
  \emph {et~al.}}]{SM_wen2013electronic}%
  \BibitemOpen
  \bibfield  {author} {\bibinfo {author} {\bibfnamefont {H.}~\bibnamefont
  {Wen}}, \bibinfo {author} {\bibfnamefont {P.}~\bibnamefont {Chen}}, \bibinfo
  {author} {\bibfnamefont {M.~P.}\ \bibnamefont {Cosgriff}}, \bibinfo {author}
  {\bibfnamefont {D.~A.}\ \bibnamefont {Walko}}, \bibinfo {author}
  {\bibfnamefont {J.~H.}\ \bibnamefont {Lee}}, \bibinfo {author} {\bibfnamefont
  {C.}~\bibnamefont {Adamo}}, \bibinfo {author} {\bibfnamefont {R.~D.}\
  \bibnamefont {Schaller}}, \bibinfo {author} {\bibfnamefont {J.~F.}\
  \bibnamefont {Ihlefeld}}, \bibinfo {author} {\bibfnamefont {E.~M.}\
  \bibnamefont {Dufresne}}, \bibinfo {author} {\bibfnamefont {D.~G.}\
  \bibnamefont {Schlom}}, \emph {et~al.},\ }\bibfield  {title} {\bibinfo
  {title} {Electronic origin of ultrafast photoinduced strain in {BiFeO$_3$}},\
  }\href {https://doi.org/https://doi.org/10.1103/PhysRevLett.110.037601}
  {\bibfield  {journal} {\bibinfo  {journal} {Phys. Rev. Lett.}\ }\textbf
  {\bibinfo {volume} {110}},\ \bibinfo {pages} {037601} (\bibinfo {year}
  {2013})}\BibitemShut {NoStop}%
\bibitem [{\citenamefont {Kundys}\ \emph {et~al.}(2010)\citenamefont {Kundys},
  \citenamefont {Viret}, \citenamefont {Colson},\ and\ \citenamefont
  {Kundys}}]{SM_kundys2010light}%
  \BibitemOpen
  \bibfield  {author} {\bibinfo {author} {\bibfnamefont {B.}~\bibnamefont
  {Kundys}}, \bibinfo {author} {\bibfnamefont {M.}~\bibnamefont {Viret}},
  \bibinfo {author} {\bibfnamefont {D.}~\bibnamefont {Colson}},\ and\ \bibinfo
  {author} {\bibfnamefont {D.~O.}\ \bibnamefont {Kundys}},\ }\bibfield  {title}
  {\bibinfo {title} {Light-induced size changes in {BiFeO$_3$} crystals},\
  }\href {https://doi.org/https://doi.org/10.1038/nmat2807} {\bibfield
  {journal} {\bibinfo  {journal} {Nat. Mater.}\ }\textbf {\bibinfo {volume}
  {9}},\ \bibinfo {pages} {803} (\bibinfo {year} {2010})}\BibitemShut {NoStop}%
\bibitem [{\citenamefont {Daranciang}\ \emph {et~al.}(2012)\citenamefont
  {Daranciang}, \citenamefont {Highland}, \citenamefont {Wen}, \citenamefont
  {Young}, \citenamefont {Brandt}, \citenamefont {Hwang}, \citenamefont
  {Vattilana}, \citenamefont {Nicoul}, \citenamefont {Quirin}, \citenamefont
  {Goodfellow} \emph {et~al.}}]{SM_daranciang2012ultrafast}%
  \BibitemOpen
  \bibfield  {author} {\bibinfo {author} {\bibfnamefont {D.}~\bibnamefont
  {Daranciang}}, \bibinfo {author} {\bibfnamefont {M.~J.}\ \bibnamefont
  {Highland}}, \bibinfo {author} {\bibfnamefont {H.}~\bibnamefont {Wen}},
  \bibinfo {author} {\bibfnamefont {S.~M.}\ \bibnamefont {Young}}, \bibinfo
  {author} {\bibfnamefont {N.~C.}\ \bibnamefont {Brandt}}, \bibinfo {author}
  {\bibfnamefont {H.~Y.}\ \bibnamefont {Hwang}}, \bibinfo {author}
  {\bibfnamefont {M.}~\bibnamefont {Vattilana}}, \bibinfo {author}
  {\bibfnamefont {M.}~\bibnamefont {Nicoul}}, \bibinfo {author} {\bibfnamefont
  {F.}~\bibnamefont {Quirin}}, \bibinfo {author} {\bibfnamefont
  {J.}~\bibnamefont {Goodfellow}}, \emph {et~al.},\ }\bibfield  {title}
  {\bibinfo {title} {Ultrafast photovoltaic response in ferroelectric
  nanolayers},\ }\href
  {https://doi.org/https://doi.org/10.1103/PhysRevLett.108.087601} {\bibfield
  {journal} {\bibinfo  {journal} {Phys. Rev. Lett.}\ }\textbf {\bibinfo
  {volume} {108}},\ \bibinfo {pages} {087601} (\bibinfo {year}
  {2012})}\BibitemShut {NoStop}%
\bibitem [{\citenamefont {Hoang}\ \emph {et~al.}(2025)\citenamefont {Hoang},
  \citenamefont {Pesquera}, \citenamefont {Hinsley}, \citenamefont {Carley},
  \citenamefont {Mercadier}, \citenamefont {Teichmann}, \citenamefont
  {Unterleutner}, \citenamefont {Knez}, \citenamefont {Dienstleder},
  \citenamefont {Ganguly} \emph {et~al.}}]{SM_hoang2025ultrafast}%
  \BibitemOpen
  \bibfield  {author} {\bibinfo {author} {\bibfnamefont {L.~P.}\ \bibnamefont
  {Hoang}}, \bibinfo {author} {\bibfnamefont {D.}~\bibnamefont {Pesquera}},
  \bibinfo {author} {\bibfnamefont {G.~N.}\ \bibnamefont {Hinsley}}, \bibinfo
  {author} {\bibfnamefont {R.}~\bibnamefont {Carley}}, \bibinfo {author}
  {\bibfnamefont {L.}~\bibnamefont {Mercadier}}, \bibinfo {author}
  {\bibfnamefont {M.}~\bibnamefont {Teichmann}}, \bibinfo {author}
  {\bibfnamefont {E.~M.}\ \bibnamefont {Unterleutner}}, \bibinfo {author}
  {\bibfnamefont {D.}~\bibnamefont {Knez}}, \bibinfo {author} {\bibfnamefont
  {M.}~\bibnamefont {Dienstleder}}, \bibinfo {author} {\bibfnamefont
  {S.}~\bibnamefont {Ganguly}}, \emph {et~al.},\ }\bibfield  {title} {\bibinfo
  {title} {Ultrafast decoupling of polarization and strain in ferroelectric
  {BaTiO$_3$}},\ }\href
  {https://doi.org/https://doi.org/10.1038/s41467-025-63045-6} {\bibfield
  {journal} {\bibinfo  {journal} {Nat. Commun.}\ }\textbf {\bibinfo {volume}
  {16}},\ \bibinfo {pages} {7966} (\bibinfo {year} {2025})}\BibitemShut
  {NoStop}%
\bibitem [{\citenamefont {Puri}\ \emph {et~al.}(2024)\citenamefont {Puri},
  \citenamefont {Patel}, \citenamefont {Cabellos}, \citenamefont
  {Rosas-Hernandez}, \citenamefont {Reynolds}, \citenamefont {Churchill},
  \citenamefont {Barraza-Lopez}, \citenamefont {Mendoza},\ and\ \citenamefont
  {Nakamura}}]{SM_puri2024substrate}%
  \BibitemOpen
  \bibfield  {author} {\bibinfo {author} {\bibfnamefont {S.}~\bibnamefont
  {Puri}}, \bibinfo {author} {\bibfnamefont {S.}~\bibnamefont {Patel}},
  \bibinfo {author} {\bibfnamefont {J.~L.}\ \bibnamefont {Cabellos}}, \bibinfo
  {author} {\bibfnamefont {L.~E.}\ \bibnamefont {Rosas-Hernandez}}, \bibinfo
  {author} {\bibfnamefont {K.}~\bibnamefont {Reynolds}}, \bibinfo {author}
  {\bibfnamefont {H.~O.}\ \bibnamefont {Churchill}}, \bibinfo {author}
  {\bibfnamefont {S.}~\bibnamefont {Barraza-Lopez}}, \bibinfo {author}
  {\bibfnamefont {B.~S.}\ \bibnamefont {Mendoza}},\ and\ \bibinfo {author}
  {\bibfnamefont {H.}~\bibnamefont {Nakamura}},\ }\bibfield  {title} {\bibinfo
  {title} {Substrate interference and strain in the second-harmonic generation
  from {MoSe}$_2$ monolayers},\ }\href
  {https://doi.org/https://doi.org/10.1021/acs.nanolett.4c03880} {\bibfield
  {journal} {\bibinfo  {journal} {Nano Lett.}\ }\textbf {\bibinfo {volume}
  {24}},\ \bibinfo {pages} {13061} (\bibinfo {year} {2024})}\BibitemShut
  {NoStop}%
\bibitem [{\citenamefont {Gonze}\ \emph {et~al.}(2009)\citenamefont {Gonze},
  \citenamefont {Amadon}, \citenamefont {Anglade}, \citenamefont {Beuken},
  \citenamefont {Bottin}, \citenamefont {Boulanger}, \citenamefont {Bruneval},
  \citenamefont {Caliste}, \citenamefont {Caracas}, \citenamefont {Cote} \emph
  {et~al.}}]{SM_gonze2009abinit}%
  \BibitemOpen
  \bibfield  {author} {\bibinfo {author} {\bibfnamefont {X.}~\bibnamefont
  {Gonze}}, \bibinfo {author} {\bibfnamefont {B.}~\bibnamefont {Amadon}},
  \bibinfo {author} {\bibfnamefont {P.}~\bibnamefont {Anglade}}, \bibinfo
  {author} {\bibfnamefont {J.-M.}\ \bibnamefont {Beuken}}, \bibinfo {author}
  {\bibfnamefont {F.}~\bibnamefont {Bottin}}, \bibinfo {author} {\bibfnamefont
  {P.}~\bibnamefont {Boulanger}}, \bibinfo {author} {\bibfnamefont
  {F.}~\bibnamefont {Bruneval}}, \bibinfo {author} {\bibfnamefont
  {D.}~\bibnamefont {Caliste}}, \bibinfo {author} {\bibfnamefont
  {R.}~\bibnamefont {Caracas}}, \bibinfo {author} {\bibfnamefont
  {M.}~\bibnamefont {Cote}}, \emph {et~al.},\ }\bibfield  {title} {\bibinfo
  {title} {Abinit: First-principles approach to material and nanosystem
  properties},\ }\href
  {https://doi.org/https://doi.org/10.1016/j.cpc.2009.07.007} {\bibfield
  {journal} {\bibinfo  {journal} {Comput. Phys. Commun.}\ }\textbf {\bibinfo
  {volume} {180}},\ \bibinfo {pages} {2582} (\bibinfo {year}
  {2009})}\BibitemShut {NoStop}%
\bibitem [{\citenamefont {Perdew}\ \emph {et~al.}(1996)\citenamefont {Perdew},
  \citenamefont {Burke},\ and\ \citenamefont
  {Ernzerhof}}]{SM_perdew1996generalized}%
  \BibitemOpen
  \bibfield  {author} {\bibinfo {author} {\bibfnamefont {J.~P.}\ \bibnamefont
  {Perdew}}, \bibinfo {author} {\bibfnamefont {K.}~\bibnamefont {Burke}},\ and\
  \bibinfo {author} {\bibfnamefont {M.}~\bibnamefont {Ernzerhof}},\ }\bibfield
  {title} {\bibinfo {title} {Generalized gradient approximation made simple},\
  }\href {https://doi.org/10.1103/PhysRevLett.77.3865} {\bibfield  {journal}
  {\bibinfo  {journal} {Phys. Rev. Lett.}\ }\textbf {\bibinfo {volume} {77}},\
  \bibinfo {pages} {3865} (\bibinfo {year} {1996})}\BibitemShut {NoStop}%
\bibitem [{\citenamefont {Bl{\"o}chl}(1994)}]{SM_blochl1994projector}%
  \BibitemOpen
  \bibfield  {author} {\bibinfo {author} {\bibfnamefont {P.~E.}\ \bibnamefont
  {Bl{\"o}chl}},\ }\bibfield  {title} {\bibinfo {title} {Projector
  augmented-wave method},\ }\href {https://doi.org/10.1103/PhysRevB.50.17953}
  {\bibfield  {journal} {\bibinfo  {journal} {Phys. Rev. B}\ }\textbf {\bibinfo
  {volume} {50}},\ \bibinfo {pages} {17953} (\bibinfo {year}
  {1994})}\BibitemShut {NoStop}%
\bibitem [{\citenamefont {Jollet}\ \emph {et~al.}(2014)\citenamefont {Jollet},
  \citenamefont {Torrent},\ and\ \citenamefont
  {Holzwarth}}]{SM_jollet2014generation}%
  \BibitemOpen
  \bibfield  {author} {\bibinfo {author} {\bibfnamefont {F.}~\bibnamefont
  {Jollet}}, \bibinfo {author} {\bibfnamefont {M.}~\bibnamefont {Torrent}},\
  and\ \bibinfo {author} {\bibfnamefont {N.}~\bibnamefont {Holzwarth}},\
  }\bibfield  {title} {\bibinfo {title} {Generation of projector augmented-wave
  atomic data: A 71 element validated table in the {XML} format},\ }\href
  {https://doi.org/https://doi.org/10.1016/j.cpc.2013.12.023} {\bibfield
  {journal} {\bibinfo  {journal} {Comput. Phys. Commun.}\ }\textbf {\bibinfo
  {volume} {185}},\ \bibinfo {pages} {1246} (\bibinfo {year}
  {2014})}\BibitemShut {NoStop}%
\bibitem [{\citenamefont {Paillard}\ \emph {et~al.}(2019)\citenamefont
  {Paillard}, \citenamefont {Torun}, \citenamefont {Wirtz}, \citenamefont
  {{\'I}{\~n}iguez},\ and\ \citenamefont
  {Bellaiche}}]{SM_paillard2019photoinduced}%
  \BibitemOpen
  \bibfield  {author} {\bibinfo {author} {\bibfnamefont {C.}~\bibnamefont
  {Paillard}}, \bibinfo {author} {\bibfnamefont {E.}~\bibnamefont {Torun}},
  \bibinfo {author} {\bibfnamefont {L.}~\bibnamefont {Wirtz}}, \bibinfo
  {author} {\bibfnamefont {J.}~\bibnamefont {{\'I}{\~n}iguez}},\ and\ \bibinfo
  {author} {\bibfnamefont {L.}~\bibnamefont {Bellaiche}},\ }\bibfield  {title}
  {\bibinfo {title} {Photoinduced phase transitions in ferroelectrics},\ }\href
  {https://doi.org/https://doi.org/10.1103/PhysRevLett.123.087601} {\bibfield
  {journal} {\bibinfo  {journal} {Phys. Rev. Lett.}\ }\textbf {\bibinfo
  {volume} {123}},\ \bibinfo {pages} {087601} (\bibinfo {year}
  {2019})}\BibitemShut {NoStop}%
\bibitem [{\citenamefont {Verstraete}\ \emph {et~al.}(2025)\citenamefont
  {Verstraete}, \citenamefont {Abreu}, \citenamefont {Allemand}, \citenamefont
  {Amadon}, \citenamefont {Antonius}, \citenamefont {Azizi}, \citenamefont
  {Baguet} \emph {et~al.}}]{SM_verstraete2025abinit}%
  \BibitemOpen
  \bibfield  {author} {\bibinfo {author} {\bibfnamefont {M.~J.}\ \bibnamefont
  {Verstraete}}, \bibinfo {author} {\bibfnamefont {J.}~\bibnamefont {Abreu}},
  \bibinfo {author} {\bibfnamefont {G.~E.}\ \bibnamefont {Allemand}}, \bibinfo
  {author} {\bibfnamefont {B.}~\bibnamefont {Amadon}}, \bibinfo {author}
  {\bibfnamefont {G.}~\bibnamefont {Antonius}}, \bibinfo {author}
  {\bibfnamefont {M.}~\bibnamefont {Azizi}}, \bibinfo {author} {\bibfnamefont
  {L.}~\bibnamefont {Baguet}}, \emph {et~al.},\ }\bibfield  {title} {\bibinfo
  {title} {Abinit 2025: New capabilities for the predictive modeling of solids
  and nanomaterials},\ }\href
  {https://doi.org/https://doi.org/10.1063/5.0288278} {\bibfield  {journal}
  {\bibinfo  {journal} {J. Chem. Phys.}\ }\textbf {\bibinfo {volume} {163}},\
  \bibinfo {pages} {164126} (\bibinfo {year} {2025})}\BibitemShut {NoStop}%
\bibitem [{\citenamefont {Kresse}\ and\ \citenamefont
  {Hafner}(1993)}]{SM_kresse1993ab}%
  \BibitemOpen
  \bibfield  {author} {\bibinfo {author} {\bibfnamefont {G.}~\bibnamefont
  {Kresse}}\ and\ \bibinfo {author} {\bibfnamefont {J.}~\bibnamefont
  {Hafner}},\ }\bibfield  {title} {\bibinfo {title} {Ab initio molecular
  dynamics for liquid metals},\ }\href
  {https://doi.org/https://doi.org/10.1103/PhysRevB.47.558} {\bibfield
  {journal} {\bibinfo  {journal} {Phys. Rev. B}\ }\textbf {\bibinfo {volume}
  {47}},\ \bibinfo {pages} {558} (\bibinfo {year} {1993})}\BibitemShut
  {NoStop}%
\bibitem [{\citenamefont {Kresse}\ and\ \citenamefont
  {Furthm{\"u}ller}(1996)}]{SM_kresse1996efficient}%
  \BibitemOpen
  \bibfield  {author} {\bibinfo {author} {\bibfnamefont {G.}~\bibnamefont
  {Kresse}}\ and\ \bibinfo {author} {\bibfnamefont {J.}~\bibnamefont
  {Furthm{\"u}ller}},\ }\bibfield  {title} {\bibinfo {title} {Efficient
  iterative schemes for ab initio total-energy calculations using a plane-wave
  basis set},\ }\href
  {https://doi.org/https://doi.org/10.1103/PhysRevB.54.11169} {\bibfield
  {journal} {\bibinfo  {journal} {Phys. Rev. B}\ }\textbf {\bibinfo {volume}
  {54}},\ \bibinfo {pages} {11169} (\bibinfo {year} {1996})}\BibitemShut
  {NoStop}%
\bibitem [{\citenamefont {Polyanskiy}(2024)}]{SM_0polyanskiy2024refractiveindex}%
  \BibitemOpen
  \bibfield  {author} {\bibinfo {author} {\bibfnamefont {M.~N.}\ \bibnamefont
  {Polyanskiy}},\ }\bibfield  {title} {\bibinfo {title} {Refractiveindex. info
  database of optical constants},\ }\href
  {https://doi.org/https://doi.org/10.1038/s41597-023-02898-2} {\bibfield
  {journal} {\bibinfo  {journal} {Sci. Data}\ }\textbf {\bibinfo {volume}
  {11}},\ \bibinfo {pages} {94} (\bibinfo {year} {2024})}\BibitemShut {NoStop}%
\bibitem [{\citenamefont {Hecht}(2012)}]{SM_0hecht2012optics}%
  \BibitemOpen
  \bibfield  {author} {\bibinfo {author} {\bibfnamefont {E.}~\bibnamefont
  {Hecht}},\ }\href@noop {} {\emph {\bibinfo {title} {Optics}}}\ (\bibinfo
  {publisher} {Pearson Education India},\ \bibinfo {year} {2012})\BibitemShut
  {NoStop}%
\bibitem [{\citenamefont {Madelung}(2004)}]{SM_madelung2004semiconductors}%
  \BibitemOpen
  \bibfield  {author} {\bibinfo {author} {\bibfnamefont {O.}~\bibnamefont
  {Madelung}},\ }\href@noop {} {\emph {\bibinfo {title} {Semiconductors: data
  handbook}}}\ (\bibinfo  {publisher} {Springer Science \& Business Media},\
  \bibinfo {year} {2004})\BibitemShut {NoStop}%
\bibitem [{\citenamefont {Ruello}\ and\ \citenamefont
  {Gusev}(2015)}]{SM_ruello2015physical}%
  \BibitemOpen
  \bibfield  {author} {\bibinfo {author} {\bibfnamefont {P.}~\bibnamefont
  {Ruello}}\ and\ \bibinfo {author} {\bibfnamefont {V.~E.}\ \bibnamefont
  {Gusev}},\ }\bibfield  {title} {\bibinfo {title} {Physical mechanisms of
  coherent acoustic phonons generation by ultrafast laser action},\ }\href
  {https://doi.org/https://doi.org/10.1016/j.ultras.2014.06.004} {\bibfield
  {journal} {\bibinfo  {journal} {Ultrasonics}\ }\textbf {\bibinfo {volume}
  {56}},\ \bibinfo {pages} {21} (\bibinfo {year} {2015})}\BibitemShut {NoStop}%
\bibitem [{ost(2020)}]{SM_osti_1197543}%
  \BibitemOpen
  \bibfield  {title} {\bibinfo {title} {Materials data on {S}n{S} by
  {M}aterials {P}roject},\ }\href@noop {} {\  (\bibinfo {year}
  {2020})}\BibitemShut {NoStop}%
\bibitem [{\citenamefont {Nye}(1985)}]{SM_nye1985physical}%
  \BibitemOpen
  \bibfield  {author} {\bibinfo {author} {\bibfnamefont {J.~F.}\ \bibnamefont
  {Nye}},\ }\href@noop {} {\emph {\bibinfo {title} {Physical properties of
  crystals: their representation by tensors and matrices}}}\ (\bibinfo
  {publisher} {Oxford university press},\ \bibinfo {year} {1985})\BibitemShut
  {NoStop}%
\bibitem [{\citenamefont {Fei}\ \emph {et~al.}(2015)\citenamefont {Fei},
  \citenamefont {Li}, \citenamefont {Li},\ and\ \citenamefont
  {Yang}}]{SM_fei2015giant}%
  \BibitemOpen
  \bibfield  {author} {\bibinfo {author} {\bibfnamefont {R.}~\bibnamefont
  {Fei}}, \bibinfo {author} {\bibfnamefont {W.}~\bibnamefont {Li}}, \bibinfo
  {author} {\bibfnamefont {J.}~\bibnamefont {Li}},\ and\ \bibinfo {author}
  {\bibfnamefont {L.}~\bibnamefont {Yang}},\ }\bibfield  {title} {\bibinfo
  {title} {Giant piezoelectricity of monolayer group {IV} monochalcogenides:
  {SnSe}, {SnS}, {GeSe}, and {GeS}},\ }\href
  {https://doi.org/https://doi.org/10.1063/1.4934750} {\bibfield  {journal}
  {\bibinfo  {journal} {Appl. Phys. Lett.}\ }\textbf {\bibinfo {volume}
  {107}},\ \bibinfo {pages} {173104} (\bibinfo {year} {2015})}\BibitemShut
  {NoStop}%
\bibitem [{\citenamefont {Gu}\ \emph {et~al.}(2002)\citenamefont {Gu},
  \citenamefont {Tani}, \citenamefont {Kono}, \citenamefont {Sakai},\ and\
  \citenamefont {Zhang}}]{SM_gu2002study}%
  \BibitemOpen
  \bibfield  {author} {\bibinfo {author} {\bibfnamefont {P.}~\bibnamefont
  {Gu}}, \bibinfo {author} {\bibfnamefont {M.}~\bibnamefont {Tani}}, \bibinfo
  {author} {\bibfnamefont {S.}~\bibnamefont {Kono}}, \bibinfo {author}
  {\bibfnamefont {K.}~\bibnamefont {Sakai}},\ and\ \bibinfo {author}
  {\bibfnamefont {X.-C.}\ \bibnamefont {Zhang}},\ }\bibfield  {title} {\bibinfo
  {title} {Study of terahertz radiation from inas and insb},\ }\href
  {https://doi.org/10.1063/1.1465507} {\bibfield  {journal} {\bibinfo
  {journal} {J. Appl. Phys.}\ }\textbf {\bibinfo {volume} {91}},\ \bibinfo
  {pages} {5533} (\bibinfo {year} {2002})}\BibitemShut {NoStop}%
\end{thebibliography}
\end{document}